\documentclass{jfm}

\usepackage[english]{babel}
\usepackage[T1]{fontenc}
\usepackage{amsmath,amssymb}
\usepackage{graphicx}
\usepackage{xcolor}
\usepackage[breaklinks,colorlinks=true, allcolors=blue]{hyperref}
\usepackage{natbib}
\usepackage{comment}
\usepackage[normalem]{ulem}

\newcommand{\bs}{\boldsymbol}
\newcommand{\mb}{\mathbb} 
\newcommand{\mc}{\mathcal} 
\newcommand{\mr}{\mathrm} 
\newcommand{\pd}{\partial}
 
\newcommand{\wh}{\widehat} 
\newcommand{\ol}{\overline}
\newcommand{\opn}{\operatorname}
 
\newcommand{\beq}{\begin{equation}}
\newcommand{\eeq}{\end{equation}}
\newcommand{\beqs}{\begin{subequations}}
\newcommand{\eeqs}{\end{subequations}}
\newcommand{\beqn}{\begin{eqnarray}}
\newcommand{\eeqn}{\end{eqnarray}}

\newcommand{\iunit}{\opn{i}}

\newcommand{\del}{\bs{\nabla}}
\newcommand{\Uvec}{\bs{\mr{U}}}
\newcommand{\uvec}{\bs{\mr{u}}}
\newcommand{\xvec}{\bs{\mr{x}}}

\newcommand{\lvec}{\bs{\mr{\ell}}}
\newcommand{\yvec}{\bs{\mr{y}}}
\newcommand{\kvec}{\bs{\mr{k}}}
\newcommand{\Xvec}{\bs{\mr{X}}}

\newcommand{\df}{\mr{d}} 
\newcommand{\dt}{\df t} 
\newcommand{\dB}{\df B_t} 
\newcommand{\dW}{\df W_t} 
\newcommand{\dF}{\df F_t} 
\newcommand{\dU}{\df \bs{\mr{U}}_t} 
\newcommand{\dx}{\df \xvec}

\newcommand{\dk}{\df \kvec}

\newcommand{\tr}{\opn{Tr}}

\newcommand{\Exp}{\mb{E}}

\newcommand{\mD}{\mc{D}} 
\newcommand{\mM}{\mc{M}} 

\newcommand{\loli}[1]{{\color{violet}}}
\newcommand{\stco}[1]{{\color{green} }}

\newcommand{\remath}[2]{\ifmmode\text{\textcolor{red}{\sout{\ensuremath{#1}}}}\else\textcolor{red}{\sout{#1}}\fi\, \textcolor{red}{#2}} 
\newcommand{\delmath}[1]{\ifmmode\text{\textcolor{red}{\sout{\ensuremath{#1}}}}\else\textcolor{red}{\sout{#1}}\fi}

\title{Simulating passive scalar advection in rough Kraichnan flows}

\author{
Long Li\aff{1,2,\corresp{\email{long.li@univ-cotedazur.fr}}},
Andr\'e L.P. Considera\aff{1,3} 
and Simon Thalabard\aff{1,2}
}

\affiliation{
\aff{1} Institut de Physique de Nice (INPHYNI), Université Côte d'Azur, CNRS -- UMR 7010, 17 rue Julien Laupr\^{e}tre, 06200, Nice, France \\ 
\aff{2} Centre Inria d'Université C\^{o}te d'Azur, 2004 Rte des Lucioles, 06560, Valbonne, France \\
\aff{3} Instituto de Matem\'{a}tica Pura e Aplicada, Estrada Dona Castorina, 110 -- Jardim Bot\^{a}nico, 22460-320, Rio de Janeiro, Brazil
} 

\begin{document}
\maketitle


\begin{abstract}
We present a systematic Eulerian study of passive scalar advection in rough two-dimensional Kraichnan flows, covering the full range of velocity roughness exponent $h\in(0,1)$. The advection--diffusion equation is integrated directly using a pseudo-spectral method at resolutions up to $2048^2$ grid points, with a frozen-noise Runge--Kutta scheme consistent with the white-in-time construction of the carrier flow. The simulations recover the duality between advected scalar and advecting flow---the smoother the carrier, the rougher the scalar---together with the predicted scaling laws for the second-order statistics. Once the scaling range is properly identified, the fourth-order flatness anomaly is measured across the whole range of $h$, in agreement with the Lagrangian estimates of Frisch \textit{et al.} (1999) and with the perturbative predictions of Bernard \textit{et al.} (1998) and Pumir \textit{et al.} (1997). Benefiting from the fact that our Eulerian simulations give direct access to the full scalar field, we also examine the probability density functions of scalar increments and the corresponding higher-order statistics, which show systematic departures from Gaussianity and from log-normality, with the strongest deviations manifesting for intermediate values of $h$. A central outcome of this work is a systematic account of how the simulation parameters, in particular the molecular diffusivity, must be adjusted with $h$, providing practical guidelines for reliable simulations; we further show that the residual deviations from the theoretical scaling laws are quantitatively accounted for by the finite spectral representation of the carrier flow. Our analysis also serve as a numerical baseline for simulating more realistic extensions of the Kraichnan model, where the carrier flow is coupled with a Gaussian multiplicative chaos and theoretical developments are limited.
\end{abstract}



\section{Introduction}\label{sec:intro}

The linear advection of a passive scalar by spatially rough velocity fields is a classical turbulence problem, which produces many of the distinctive signatures of turbulence, from gradient amplification and cascade processes to scaling and dissipative anomalies \citep{shraiman2000scalar,gotoh2013passive,benzi2023lectures}.  In a forced-dissipative setting, one considers the advection-diffusion equation
\beq\label{eq:kraichnan}
    \pd_t \theta + \uvec \cdot \del \theta 
    = 
    \kappa \Delta \theta + f, 
\eeq
describing the transport of a scalar quantity $\theta$ such as temperature, passively advected by a turbulent fluid  in a  $d-$dimensional ambient Euclidean space. Here, $\uvec$ is the advecting velocity field, $\kappa > 0$ is the molecular diffusivity, and $f$ is an external forcing term. The qualification ``passive'' refers to the fact that $\theta$ is carried by the velocity field without influencing the flow itself. While in principle, one would wish to consider linear advection by a velocity field $\uvec$ solving the Navier-Stokes equations, the Kraichnan model provides a crude yet fruitful simplification, which was initially introduced by \citet{kraichnan1968small,kraichnan1974convection,kraichnan1994anomalous} and later popularized in particular by a series of work from Gawedzki and collaborators---see \citep{falkovich2001particles,gawedzki2008stochastic} and references therein. One takes the vector field $\uvec$ to be a centered homogeneous Gaussian random field, white in time and only H\"older continuous in space, prescribed by the  covariance
\beq\label{eq:velocity_covariance}
    \Exp [ u_i(t,\xvec)\, u_j(s,\yvec) ] 
    = 
    \delta(t-s)\, \mD_{ij}(\xvec-\yvec), \quad 
    1 \le i,j \le d,
\eeq
where $\mD_{ij}(\xvec-\yvec)$ decays $\propto |\xvec-\yvec|^{2h}$, $h \in (0,1)$ for separations in the inertial range and $\sum_{i=1}^d \pd_i \mD_{ij} = 0$ to ensure incompressibility. The scaling of the spatial correlation tensor $\mD_{ij}$ encodes the spatial regularity of the velocity field, prescribing a field that is almost surely H\"older continuous with exponent $h'$ for every $h' < h$. Hereafter, we abuse the terminology slightly and call $h$ the H\"older exponent of the velocity field. Naturally, the assumptions of temporal decorrelation and Gaussian statistics are idealizations of turbulent velocity fields, which generically exhibit temporal correlations and non-Gaussian fluctuations arising from intermittency. These assumptions, however, lead to explicit analytical  closure of correlation functions for $\theta$ in the form of a Hopf hierarchy. This additional  degree of analytical tractability makes the Kraichan model a useful caricature or turbulent transport, which led to fundamental concepts such as zero mode, statistical conservation laws \citep{bernard1998slow,benzi2023lectures,thalabard2024zero} and spontaneous stochasticity \citep{eyink2025beyond}.

One of the most remarkable features of the Kraichnan model is that the scalar field $\theta$ exhibits intermittent statistics even though the driving velocity field is Gaussian. Broadly defined, intermittency refers to the breakdown of self-similar behavior for multipoint statistics, typically signaled by anomalous (multifractal) scaling and increasing departures from Gaussianity as one moves down to smaller and smaller scales. In turbulence studies, a standard diagnostic is the flatness anomaly, \emph{e.g.} scale dependency for the flatness of the scalar increments $\Delta \theta = \theta(\xvec + \lvec) - \theta(\xvec)$ \citep{chevillard2015peinture,benzi2023lectures}. More concretely, denoting the scalar structure functions by
\beq\label{eq:structure_functions}
    S_p^\theta(\ell) 
    = 
    \Exp \left[ |\theta(\xvec + \lvec) - \theta(\xvec)|^p \right] 
    \propto 
    \ell^{\zeta_p^\theta}, \quad 
    |\lvec| = \ell,
\eeq
the flatness is defined as the ratio 
\beq\label{eq:flatness}
    F_4^\theta(\ell) 
    = 
    \frac{S_4^\theta(\ell)}{(S_2^\theta(\ell))^2} \propto \ell^{-\Delta_4(h)}, \quad
    \Delta_4(h)
    =
    2\zeta_2^\theta - \zeta_4^\theta.
\eeq
A positive flatness exponent $\Delta_4(h) > 0$ indicates a non-trivial scale-dependency for the fourth-order statistics, implying that scalar increments become increasingly non-Gaussian and heavy tailed at small scales.  

In the Kraichnan model, numerical estimates for the flatness exponent $\Delta_4(h)$ were obtained in the late 90's via Lagrangian Monte-Carlo methods \citep{vergassola1997structures,frisch1999lagrangian,mazzino2000passive}, where the moments of the scalar are estimated from stochastic particle trajectories using Feynman-Kac representations. The advection-diffusion equation \eqref{eq:kraichnan} is bypassed entirely and the passive scalar field itself is not simulated---instead replaced by Cholesky factorisation of $N$-point correlators. On the other hand, Eulerian simulations apply a spatio-temporal discretization of \eqref{eq:kraichnan} with suitable care for the underlying stochastic calculus, evolving $\theta$ directly on a grid. Standard references are \citet{Chen1998PF} and \citet{fairhall1997direct}, which consider two-dimensional advection---we are unaware of more recent studies. Both studies tend to overestimate the scaling exponents for second-order statistics when $h> 1/2$, and underestimate them when $h<1/2$, respectively. The former  does not report results for the flatness exponent, while the latter suggests $\Delta_4(h) \simeq 0.5$ for $h \simeq 0.5$. To this date the Lagrangian simulations of \citet{frisch1999lagrangian} therefore remain the main numerical benchmark for the signature of intermittency in scalar statistics. Analytical predictions for $\Delta_4(h)$ are available from perturbative expansions around solvable Gaussian limits. These include the cases $h \to 0$ \citep{bernard1998slow}, $h \to 1$ \citep{pumir1997perturbation} in any dimension, together with the limit $d \to \infty$ \citep{chertkov1996anomalous}. The numerical estimates obtained from the Lagrangian schemes are consistent with these perturbative calculations \citep{vergassola1997structures,mazzino2000passive}. As an aside, the flatness anomaly in the Kraichnan model is of the same order of magnitude as that observed for passive scalar advection in a 3D homogeneous incompressible Navier-Stokes velocity field; the classical Kolmogorovian phenomenology prescribes $h=1/3$ for the velocity and direct numerical simulations suggest $\Delta_4(h) \simeq  0.35$ for the scalar \citep{watanabe2004statistics,gotoh2015power,iyer2018steep}.

In this work, we revisit the Kraichnan model from the numerical point of view and perform a systematic study of the Kraichnan passive scalar in steady-state, focusing on the intermittency of $\theta$. We work on the two-dimensional torus and use the classical pseudo-spectral method for the spatial discretization; the time integration is carried out by explicit Runge--Kutta schemes. Our Eulerian simulations successfully recover the Lagrangian results of \citet{frisch1999lagrangian} once the appropriate inertial scaling ranges are identified. For larger values of $h$, finite-size effects dominate, which progressively erase the inertial-range scaling. These effects come from the presence of numerical dissipation and from the slow convergence of random Fourier series describing the scalar. We however display  a finite-size scaling prediction that accounts for resolution constraints and agrees well with the numerical data across the full range of $h$. Beyond the scaling exponents, our simulations provide access the full statistics of the scalar increments, revealing deviations from log-normal scaling. These deviations vary non-monotonically with $h$ and are most pronounced for $h \simeq 1/2$.

The motivation for yet another numerical study of the Kraichnan model is twofold. First, the early Eulerian simulations of \citet{Chen1998PF} and \citet{fairhall1997direct} are largely confined to a few values of $h$ and do not span the allowed range $(0,1)$ in a satisfactory manner; in this work, we perform extensive simulations covering the full range. A central outcome of this study is the systematic identification of how simulation parameters, such as the molecular diffusivity $\kappa$, must be adjusted as a function of $h$ in order to obtain reliable results, from second-order statistics to the flatness exponent $\Delta_4(h)$. This analysis is summarized in the form of a practical table (see Table~\ref{tab:h_params}), providing guidelines for the choice of parameters needed to produce accurate simulations at each value of $h$. Second, and more forward-looking, we aim to establish the capabilities and limitations of Eulerian Kraichnan simulations, with a view toward extensions of the Kraichnan model for which Lagrangian methods are not readily available or where zero-mode theory does not provide perturbative predictions. Let us elaborate on this point. As mentioned earlier, real turbulent flows are not Gaussian, exhibiting signatures of intermittency at the level of the carrier flow itself. In recent years, considerable effort has been devoted to the construction of synthetic random velocity fields that incorporate this intermittent, multifractal structure observed in real turbulent flows. This program was initiated by \citet{robert2008hydrodynamic} and has been developed extensively by Chevillard and collaborators \citep{chevillard2010stochastic,chevillard2015peinture,chevillard2019skewed, pereira2016dissipative,chatelain2026spatio}. The central idea is to obtain more realistic velocity ensembles by introducing intermittency directly into the velocity field, typically by coupling a Gaussian component to a Gaussian multiplicative chaos \citep{rhodes2014gaussian}---this provides a stochastic representation of Kolmogorov's refined similarity theory \citep{kolmogorov1962refinement,ruffenach2026spatio}.

So far, however, the literature on these multifractal extensions has focused either on the numerical analysis of the random velocity field itself, or, when transport is considered, on the motion of Lagrangian particles rather than scalar fields \citep{reneuve2020flow,considera2023spontaneous}. Theoretical developments for these extensions remain limited \citep{considera2026transport}, and for passive scalar transport they are, to the best of our knowledge, nonexistent. A natural first step is therefore to conduct a numerical study of the transport of passive scalar fields by intermittent, non-Gaussian flows, addressing the following intriguing question: how does intermittency in the velocity field affect the intermittency of the passive scalar? This question will be the subject of a forthcoming companion paper. Since no theoretical predictions are available in the multifractal setting, Eulerian simulations are the only tool at hand, and a reliable numerical baseline for the classical Kraichnan model becomes an essential prerequisite. The present work serves precisely this purpose, and can be seen as the groundwork for the larger program detailed above.

The remainder of the paper is organized as follows. Section~\ref{sec:background} introduces the Kraichnan model and summarizes the theoretical predictions. Section~\ref{sec:setup} describes the pseudo-spectral numerical framework and the computational setup. Section~\ref{sec:second_order} presents the second-order statistics of the random flow and passive scalar. Section~\ref{sec:intermittency} investigates scalar intermittency through higher-order statistics. Section~\ref{sec:finitesize} examines the main finite-size effects of the numerical approximation. Finally, Section~\ref{sec:conclu} summarizes the main findings and discusses possible directions for future work.


\section{Theoretical background}\label{sec:background}

This section recalls a few basic facts about the Kraichnan model, from its formulation as a stochastic partial differential equation (SPDE) under either It\^o or Stratonovich stochastic calculus, to Hopf hierarchy and scaling exponents estimates.

\subsection{SPDE formulation}

\subsubsection{Stratonovich-Kraichnan}

The advection-diffusion equation \eqref{eq:kraichnan} describes the transport of the scalar field $\theta$ by an incompressible velocity field $\uvec$ subject to large-scale stirring and molecular diffusion. With $\uvec$ being random, the transport term $\uvec \cdot \del \theta$ acts as a multiplicative noise which requires careful stochastic interpretation. In stochastic differential form, the simplest (correct) formulation of the Kraichnan model reads
\beq\label{eq:kraichnan_strat}
    \df \theta_t + \circ \dU \cdot \del \theta_t
    =
    \kappa \Delta \theta_t\, \dt + \dF,
\eeq
where $\circ$ denotes Stratonovich integration. The transport term in the Stratonovich SPDE  \eqref{eq:kraichnan_strat} involves the random flow increment $\dU$, which is white in time and correlated in space, while $\dF$ is an independent Gaussian scalar forcing. The forcing is specified in detail below in \S~\ref{sec:ito-kraichnan}. Comparing Eqs.~\eqref{eq:kraichnan_strat} and \eqref{eq:kraichnan}, we have the formal identifications $\uvec(\xvec,t) \equiv \dU(\xvec)/\dt$ and $f(\xvec,t)  \equiv \dF(\xvec)/\dt$. The Stratonovich formulation is natural when the white-in-time model is viewed as the short-correlation-time limit of smooth random flows \citep[see, \emph{e.g.},][]{eugene1965relation,flandoli2021transport,flandoli2022additive}. Let us however observe that, strictly speaking, the process $\Uvec_t$  is not a velocity field: dimensional analysis shows that it has the units of a displacement field, and its spatial correlation kernel scales as a diffusivity rather than a velocity. We refer to $\Uvec_t$ as the \emph{generalized velocity field}---following the lecture notes of \citet{gawedzki2006simple}. 

\subsubsection{The generalized velocity field}\label{ssec:generalized}

We work on full two-dimensional space $\mb{R}^2$ and take the generalized velocity field to be spatially homogeneous and incompressible. Following \citet{chaves2003lagrangian}, the field is most easily represented in Fourier variables, \emph{e.g.} 
\beq\label{eq:flow_fourier}
    \dU (\xvec)
    =
    \sqrt{\dfrac{D_0}{Z}} \int_{\mb{R}^2} \dk\;
    e^{\iunit \kvec\cdot\xvec}\ \dfrac{g_\eta(|\kvec|)}{(|\kvec|^2+1)^{1/2+h/2}}\,
    \wh{\kvec}_\perp \df {B}_t(\kvec),
\eeq
where  
$\wh{\kvec}_\perp  = (-k_y/|\kvec|,k_x/|\kvec|)$. The processes $\{B_t(\kvec)\}$ are complex Brownian motions satisfying the Hermitian symmetry 
\beq\label{eq:hermitian}
B_t (-\kvec) = \ol{B_t (\kvec)},
\eeq
so that the generalized velocity field has real-valued components, with  their covariance  explicitly given by
\beq
    \Exp \left[ B_s (\kvec_1) \ol{B_t} (\kvec_2) \right]
    = 
    \min(s,t)\, \delta (\kvec_1 - \kvec_2),
\eeq
and with $\dB(\kvec)$ denoting the corresponding stochastic time differential. $g_\eta$ is a low-pass filter acting as a small-scale regularization, such that $g_\eta(k) \to 1$ (resp. $ g_\eta(k) \to 0$) for $k\ll 1/\eta$  (resp. $ k \gg 1/\eta$). The parameter $h\in(0,1)$ is the spatial H\"older exponent of the random flow in the limit $\eta \to 0$. The coefficients $D_0$ and $Z=Z(\eta,h)$ fix the amplitude of the spatial covariance at equal time. Specifically, the generalized field is specified by the covariance 
\beqs\label{eq:flow_covariance}
\beq
    \Exp \left[ \df U_t^l (\xvec) \df U_s^m (\xvec+\lvec) \right]
    =
    \delta(t-s)\, \dt\, \mD^{lm} (\lvec),
\eeq
\beq
    \mD^{lm} (\lvec)
    =
    \dfrac{D_0}{Z} \int_{\kvec \in \mb{R}^2} \dk\; 
    e^{\iunit\kvec\cdot\lvec} \frac{g_\eta^2(|\kvec|)}{(|\kvec|^2+1)^{1+h}}P^{lm}(\kvec)
\eeq
\eeqs
with $l,m \in \{x,y\}$ and $P^{lm}(\kvec) = \delta^{lm} - \frac{k^l k^m}{|\kvec|^2}$ the (Leray) projector on two-dimensional divergence-free fields. Eq.~\eqref{eq:flow_covariance} prescribes the flow to be statistically isotropic. In particular, the two-point vectorial correlator depends on $ \ell=|\lvec|$ only as
\beqs\label{eq:flow_trace_covariance}
\beq
    \Exp \left[ \dU (\xvec) \cdot \dU (\xvec+\lvec) \right]
    =
    \dt\, \tr \mD (|\lvec|),
\eeq
\beq\label{eq:flow_trace_covariance_b}
    \tr \mD (\ell)  
    =
    \dfrac{2\pi D_0}{Z} \int_{\mathbb R_+} \df k\; 
    J_0 (k \ell) \frac{k\, g_\eta^2(k)}{(k^2+1)^{1+h}},
\eeq
\eeqs
involving the zero-order Bessel function $J_0$. The normalization factor Z is taken such that     $\tr \mD (0) = 2 D_0$, implying $\Exp \left[ |\dU|^2 \right] = 2 D_0$ and $\mD^{xx}(0) = \mD^{yy}(0) = D_0$. Note that $\mD^{xy}(0) = \mD^{yx}(0) = 0$.

The factor  $k^2+1$  in the denominator inside the integrand of Eq.~\eqref{eq:flow_trace_covariance_b} acts as a large-scale cutoff on scale $O(1)$ and disappears in the large-$k$ limit. The precise shape of the cutoff is an arbitrary choice. Binning the trace over shells of radius yields the isotropic power spectrum
\beq\label{eq:powerspectrum}
    E_{\Uvec}(k) 
    \sim 
    \dfrac{2\pi D_0}{Z} \frac{g_\eta^2(k)}{k^{1+2h}}, \quad 
    k \to \infty.  
\eeq
This asymptotic regime prescribes the small separation asymptotics in the joint limit $\eta,\ell \to 0$. The order of those limits matters---see Appendix~\label{sec:tauberian} for computational details. Taking the limit $\eta \to 0$ first yields the rough decay rates
\beq\label{eq:roughdecay}
    \tr \mD (\ell) 
    \approx 
    2 D_0 -  2 D_1 \ell^{2h}, \quad 
    D_1
    := 
    - \dfrac{D_0\Gamma(-h)}{2^{2h}\, \Gamma(h)},
\eeq
which are independent of the regularization scheme, \emph{e.g.} of $g_\eta$. The rough asymptotics represents the inertial range scaling and requires the separation of scales  $\eta \ll \ell \ll 1$. In particular, the classical Kolmogorovian (K41) turbulence \citep{kolmogorov1991local,chevillard2019skewed} regime corresponds to the case $h=1/3$.

\begin{figure}
    \centering
    \includegraphics[width=0.5\textwidth]{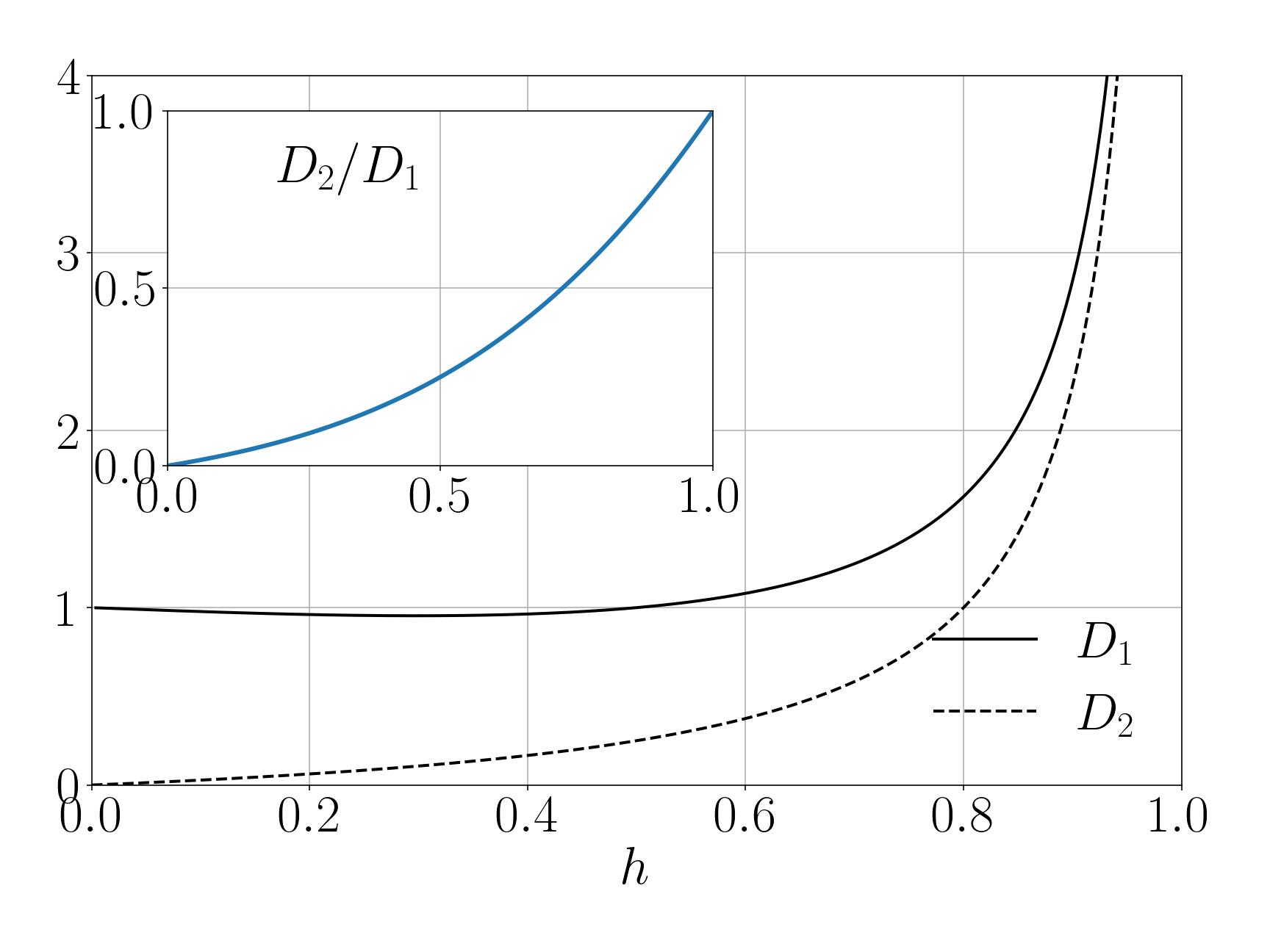}
    \caption{Coefficients $D_1,D_2$ featured in Eqs.~\eqref{eq:roughdecay}--\eqref{eq:smoothdecay}, taking $D_0=1$. The inset shows the ratio $D_2/D_1$.}
    \label{fig:D12}
\end{figure}

On the other hand, taking the limit $\ell \to 0$ first yields the Batchelor regime
\beq\label{eq:smoothdecay}
    \tr \mD (\ell) 
    \approx 
    2 D_0 - 2 D_2 k_\eta^{2-2h} \ell^2,  
\eeq 
where $D_2$ is a non-universal coefficient depending on the choice of cutoff function $g_\eta$.  For example, 
\beq\label{eq:D2}
    D_2 
    \sim 
    \begin{cases}
        & \dfrac{D_0h}{4(1-h)} \quad \text{for the sharp cutoff} g_\eta(k) := 1_{k<k_\eta} \\
	& \dfrac{D_0h}{4\Gamma(1-h)} \quad \text{for the Gaussian cutoff} g_\eta(k) := \exp \left( -\frac{k^2}{2 k_\eta^2} \right)
    \end{cases}.
\eeq
The smooth rate of decay requires the separation of scales $ \ell \ll \eta \ll 1$, which can be interpreted as representative of  a dissipative region. For the reader's convenience, Fig.~\ref{fig:D12} shows the coefficients $D_1, D_2$ as a function of $h$ for the case of a sharp cutoff. They both diverge as $h\to 1$ with $D_1/D_2 = O(1)$, while $D_2 = o(D_1)$ for $ h \to 0$. 

\begin{figure}
    \includegraphics[width=0.32\textwidth]{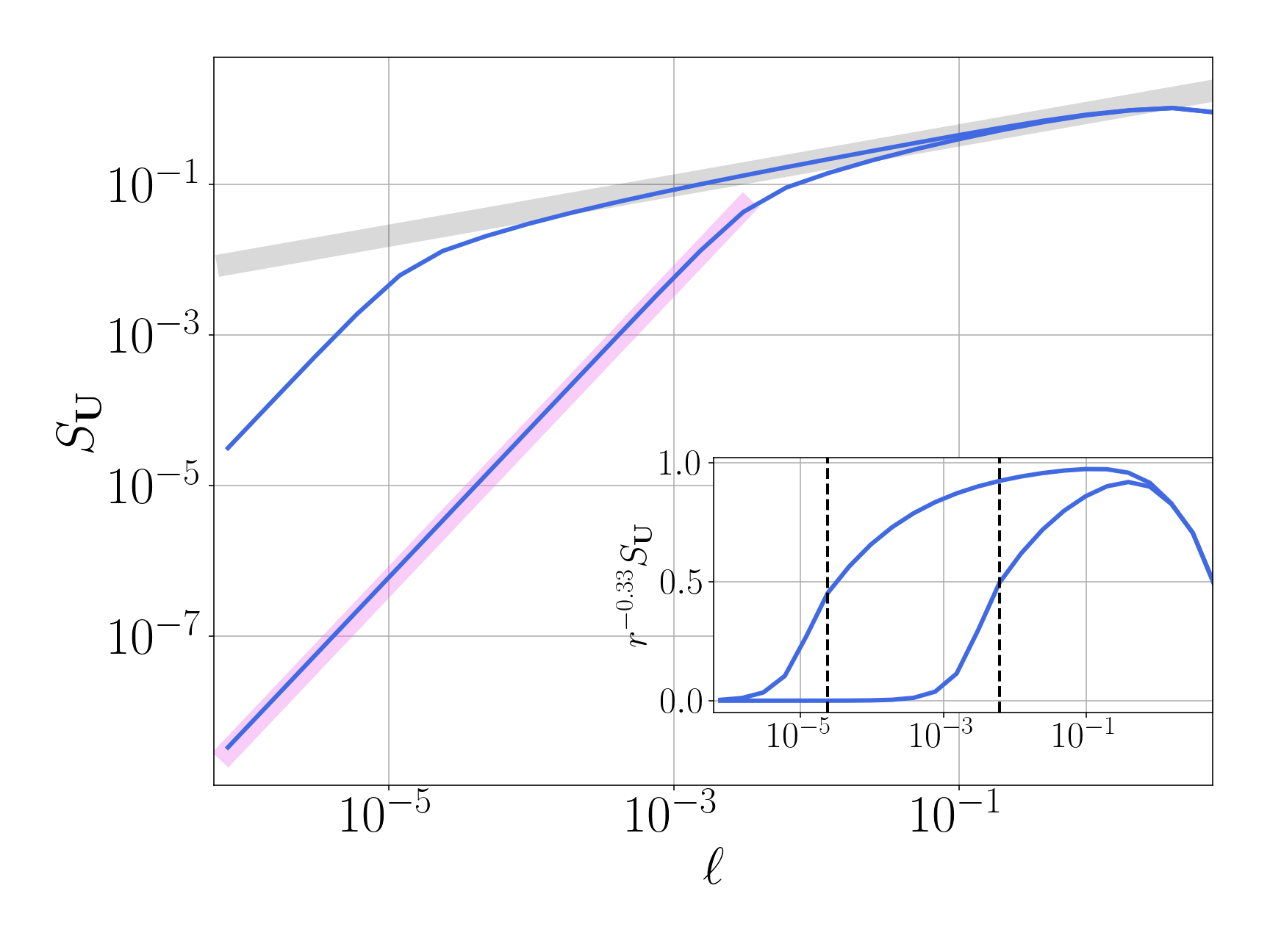}
    \includegraphics[width=0.32\textwidth]{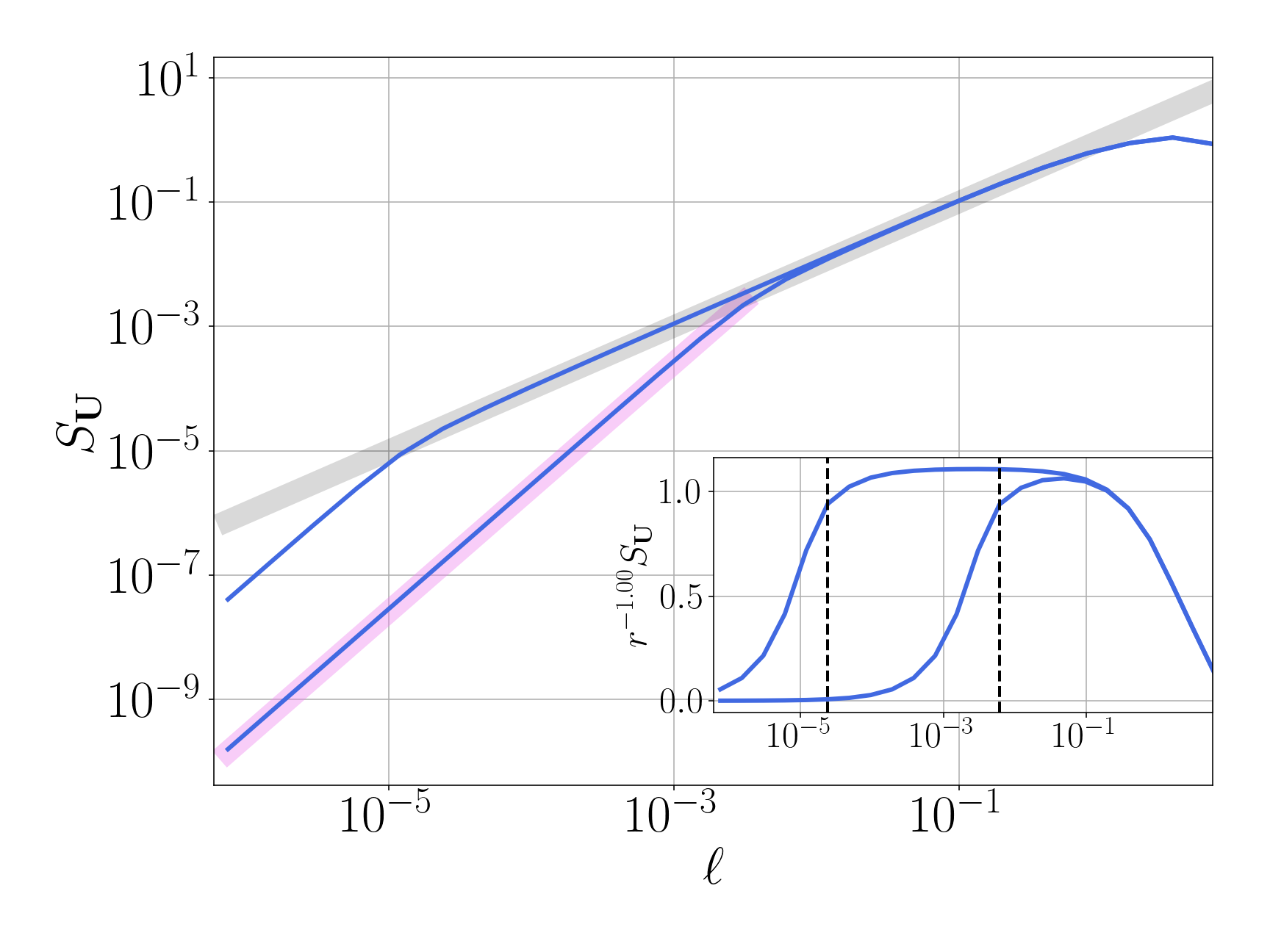}
    \includegraphics[width=0.32\textwidth]{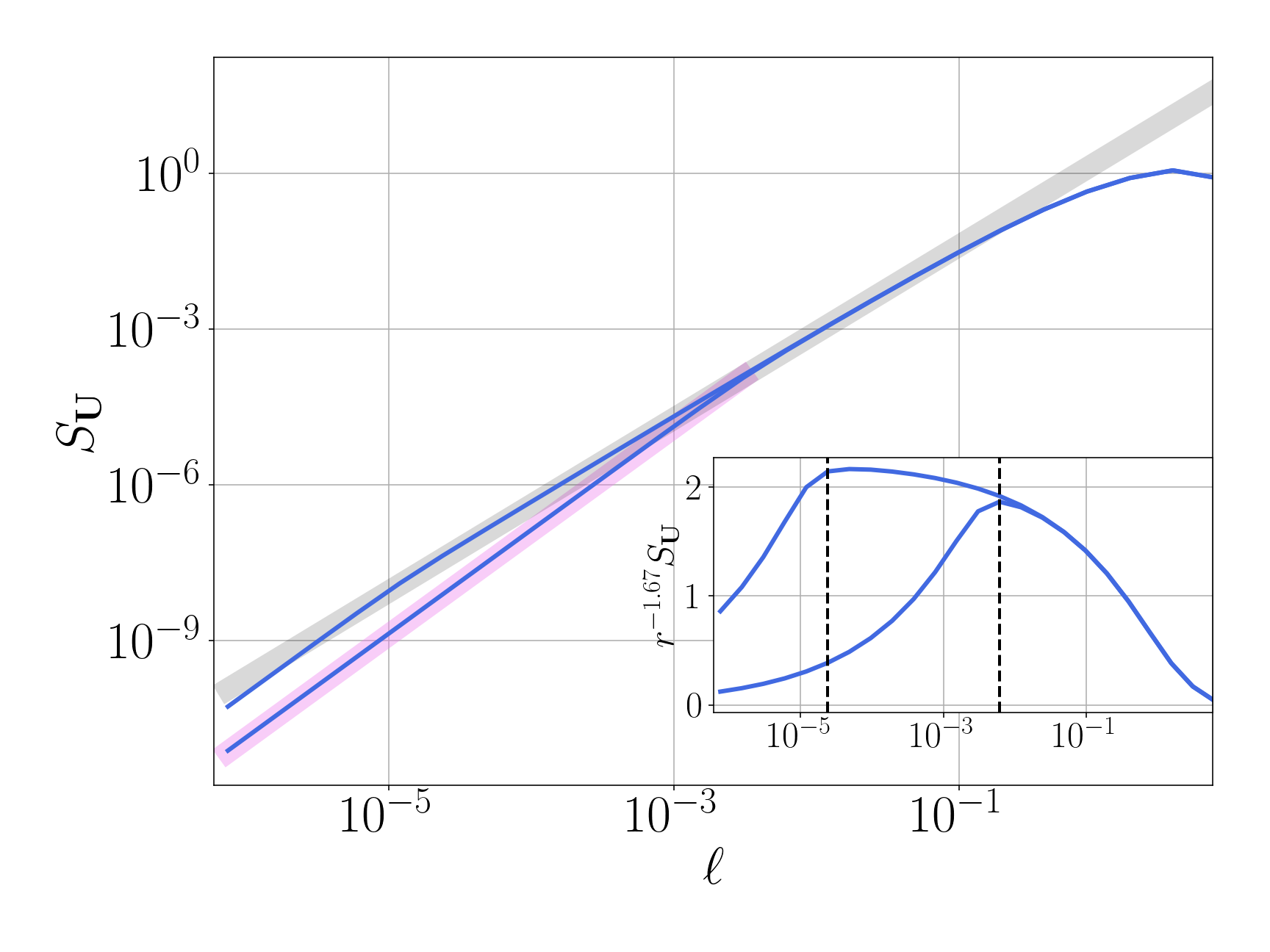}
    \caption{Numerical estimates of the scaling part $S_{U} (\ell) = 2 D_0 - \tr \mD (\ell)$ obtained using a Riemann discretization of Eq.~\eqref{eq:flow_trace_covariance} over integer wave numbers with the step cutoff $g_\eta = 1_{k<k_\eta}$, where $k_\eta = 1024$ and  $k_\eta=262144$. The thick grey and red lines indicate the predicted scaling laws $2 D_1 \ell^{2h}$ and $2 D_2 k_\eta^{2-2h} \ell^2$, respectively. The insets show the data compensated by the rough scaling $\ell^{2h}$ and the vertical lines mark $ \ell = \eta$. The three panels correspond to $h=1/6$ (left), $h=1/2$ (middle) and $h=5/6$ (right).}
    \label{fig:inertial}
\end{figure}

For finite $\eta$, both the inertial and dissipative scaling regions required for Eq.~\eqref{eq:roughdecay} and Eq.~\eqref{eq:smoothdecay} are obscured by finite size effects, whose importance depends on the value of the H\"older exponent $h$. Fig.~\ref{fig:inertial} shows numerical estimates of the scaling contributions $S_U (\ell) := 2 D_0 - \tr \mD (\ell)$ estimated using a Riemann discretization of Eq.~\eqref{eq:flow_trace_covariance} over integer wave numbers with the step cutoff $g_\eta = 1_{k<k_\eta}$. The rough scaling regime $\propto \ell^{2h}$ is most clearly observed for $h \simeq 1/2$. As $h$ decreases towards $0$, the inertial range gets narrower, and is seen over the effective scaling region $\eta \ll \ell \lesssim 1$. For example, one must resolve the rather prohibitive cutoff $k_\eta = 262144$ to observe even one decade of scaling for $h=1/6$. A  similar narrowing of the inertial range is seen for $h \to 1$, except that it instead develops over the effective region $\eta \lesssim \ell \ll 1$.
 
To conclude, we point out that the covariance structure decomposes into longitudinal and transverse components as 
\beqs\label{eq:flow_covariance_par}
\beq
    \mD (\lvec) 
    =
    \mD^\parallel (\ell) \hat \lvec \otimes \hat \lvec + \mD^\perp (\ell) \hat \lvec_\perp \otimes \hat \lvec_\perp,
\eeq
\beq
    \mD^\parallel (\ell) 
    = 
    \dfrac{\pi D_0}{Z} \int_{\mb{R}_+} \df k\;
    \bigl( J_0 (k \ell) + J_2 (k \ell) \bigr) \frac{k\, g_\eta^2 (k)}{(k^2+1)^{1+h}},
\eeq
\beq 
    \mD^\perp (\ell) 
    = 
    \dfrac{\pi D_0}{Z} \int_{\mathbb R_+} \df k\;
    \bigl( J_0 (k \ell) - J_2 (k \ell) \bigr) \frac{k\, g_\eta^2 (k)}{(k^2+1)^{1+h}},
\eeq
\eeqs
Each contribution admits rough and smooth asymptotic behavior, \emph{e.g.} $\mD^{\parallel} (\ell) \approx D_0 - D_1^{\parallel} \ell^{2h}$, etc. The corresponding scaling coefficients, with notations paralleling Eqs.~\eqref{eq:roughdecay}--\eqref{eq:smoothdecay} are
\beq\label{eq:coeff}
    D_1^{\parallel} = \dfrac{D_1}{1+h}, 
    \quad  
    D_1^{\perp} = \dfrac{(1+2h)D_1}{1+h},
    \quad 
    D_2^{\parallel} = \dfrac{D_2}{2}, 
    \quad  
    D_2^{\perp} = \dfrac{3 D_2}{2}.
\eeq
Note that in coordinate form, Eq.~\eqref{eq:flow_covariance_par} implies $\mD_{ij} (\lvec) = \left( \mD^\parallel (\ell) - \mD^\perp (\ell) \right) \ell_i \ell_j / \ell^2 + \mD^\perp \delta_{ij}$ for $i,j \in \{x,y\}$. The longitudinal and transverse correlators then determine the corresponding structure functions, \emph{e.g.} $S_2^\parallel = 2 \left( D^\parallel(0) - D^\parallel(\ell) \right)$, and $S_2^\perp = 2 \left( D^\perp (0) - D^\perp (\ell) \right)$.

\subsubsection{It\^o-Kraichnan}\label{sec:ito-kraichnan}

It proves convenient to reformulate the Kraichnan transport \eqref{eq:kraichnan_strat} in its equivalent It\^o form. Let us first specify the scalar forcing. Common choices in the turbulence literature include random forcing or gradient forcing \citep{gotoh2015power}. We here restrict our analysis to the former case, and prescribe the forcing to be Gaussian, white in time, spatially homogeneous and isotropic, and independent from the generalized velocity field:
\beq\label{eq:forcing}
    \dF(\xvec)
    =
    \int_{\mb{R}^2} \dk\;
    e^{\iunit\kvec\cdot\xvec} \phi(\kvec)\, \dW(\kvec),
\eeq
where the processes $\{W_t(\kvec)\}$ are complex Brownian motions satisfying the Hermitian symmetry \eqref{eq:hermitian} and independent from the processes $B_t(\kvec)$ entering the generalized field \eqref{eq:flow_fourier}. We take the Fourier kernel $\phi$ concentrated at small wavenumbers, so that the scalar variance is injected at large scales, \emph{e.g.}
\beq\label{eq:forcing_kernel}
    \phi(\kvec)
    =
    \sqrt{\dfrac{\varepsilon_\theta}{\pi k_f^2}} 
    \exp \left[ -\frac{1}{2} \left( \frac{|\kvec|}{k_f} \right)^2 \right],
\eeq
where $k_f$ represents the forcing scale, $\varepsilon_\theta$ the injection rate of scalar variance. In physical space, Eqs.~\eqref{eq:forcing} and~\eqref{eq:forcing_kernel} imply
that the forcing has covariance
\beq\label{eq:forcing_covariance}
    \Exp \left[ \dF(\xvec) \df F_s(\xvec+\lvec) \right]
    =
    \delta(t-s)\, \dt\, \Phi(|\lvec|), \quad
    \Phi(\ell)
    =  
    \varepsilon_\theta \exp \left( -\dfrac{k_f^2\ell^2}{4} \right).
\eeq
The It\^{o} version of the Stratonovich-Kraichnan transport~\eqref{eq:kraichnan_strat} driven by the generalized field \eqref{eq:flow_fourier} under the scalar forcing \eqref{eq:forcing} is 
\citep[see, e.g.,][]{rowan2024anomalous,drivas2025anomalous}
\beq\label{eq:kraichnan_ito_iso}
    \df \theta_t + \dU\cdot\del\theta_t
    =
    \left( \kappa + \frac{D_0}{2} \right) \Delta\theta_t\, \dt + \dF,
\eeq
where we recall the definition $D_0 = \tr \mD(0)/2$ from Eq.~\eqref{eq:flow_trace_covariance} for the diffusion coefficient. Using It\^{o} calculus, one verifies that the It\^{o} drift term $\left(\frac{D_0}{2}\right) \Delta\theta_t\, \dt$ ensures, in the absence of diffusivity and forcing,  that the scalar is transported along the Lagrangian trajectories, \emph{e.g.} $\theta_t (\Xvec_t) = \theta_0 (\Xvec_0)$ almost surely for random trajectories satisfying $\df \Xvec_t = \dU(\Xvec_t)$. While the Stratonovich formulation of Eq.~\eqref{eq:kraichnan_strat} and the It\^o formulation of Eq.~\eqref{eq:kraichnan_ito_iso} are mathematically equivalent, their respective merits serve different purposes. The Stratonovich formulation will prove useful for numerical simulations, while the It\^o proves convenient for theoretical developments.

\subsection{Statistical properties}

\subsubsection{Hopf hierarchy}\label{sec:hopf}

The Kraichnan model yields a recursive Hopf hierarchy for the $n$-point scalar correlation functions 
\beq\label{eq:Cn}
    C_n (\xvec_1,\ldots,\xvec_n,t)
    =
    \Exp \left[ \theta_t(\xvec_1) \cdots \theta_t(\xvec_n) \right].
\eeq
The hierarchy can be derived from stochastic calculus starting from the It\^o formulation in Eq.~\eqref{eq:kraichnan_ito_iso}. In full generality, it prescribes the order-by-order relations
\beq\label{eq:dCndt}
    \dfrac{d}{dt} C_n (\xvec_1, \cdots,\xvec_n,t) 
    = 
    \mM_n C_n + \sum_{1 \le i<j \le n} \Phi(|\lvec_{ij}|) \wh{C}^{ij}_{n-2}, \quad 
    \lvec_{ij} 
    := 
    \xvec_j - \xvec_i,
\eeq
where the $n$-particle propagator $\mM_n$ is given by 
\beq
    \mM_n
    :=
    \kappa \sum_{i=1}^n \Delta_{i} + \dfrac{D_0}{2} \left( \sum_{i=1}^n \nabla_i \right)^2 
    - \sum_{1 \le i<j \le n} \left( \mD(0) - \mD(\lvec_{ij}) \right) : \nabla_i \otimes \nabla_j, 
\eeq
and $\wh{C}^{ij}_{n-2}$ denotes the reduced $(n-2)$-point scalar correlation function obtained by omitting the fields at $\xvec_i$ and $\xvec_j$ in Eq.~\eqref{eq:Cn}, \emph{e.g.} $\wh{C}^{ij}_{n-2} = \Exp\left[ \prod_{\substack{k=1\\ k\neq i,j}}^n \theta_t(\xvec_k) \right]$. The subscript $i$ indicates that the differential operator acts on the variable $\xvec_i$, \emph{e.g.} $\Delta_1 = \pd^2_{x_1x_1} + \pd^2_{y_1y_1}$ and $:$ denotes Frobenius inner product. Note that in coordinate form, the operator $\mM_n$ reads
\begin{align}
    \mM_n
    &=
    \left( \kappa + \dfrac{D_0}{2} \right) \sum_{i=1}^n \left( \pd^2_{x_ix_i} + \pd^2_{y_iy_i} \right)
    + D_0 \sum_{1 \le i<j \le n} \left( \pd^2_{x_ix_j} + \pd^2_{y_iy_j} \right) \nonumber\\
    &-
    \sum_{1 \le i<j \le n} \sum_{l,m \in \{x,y\}} \left( \mD_{lm}(0) - \mD_{lm}(\lvec_{ij}) \right)\pd^2_{l_i m_j}.
\end{align}
Eq.~\eqref{eq:dCndt} combined with the asymptotic relations of \S~\ref{ssec:generalized} determines the asymptotic behavior of the scalar correlation functions and of the structure functions defined in Eq.~\eqref{eq:structure_functions}.

\subsubsection{Two-point correlations and dimensional scaling}\label{sec:twopoint}

Taking $n=2$ in the Hopf hierarchy of Eq.~\eqref{eq:dCndt} yields
\beq\label{eq:C2}
    \dfrac{\df}{\dt} C_2 (\xvec_1,\xvec_2,t)
    = 
    \mM_2\, C_2 + \Phi\left(|\lvec|\right), \quad
    \lvec
    :=
    \xvec_2 - \xvec_1,
\eeq
with the two-particle propagator $\mM_2$ taking the form
\beq
    \mM_2
    =
    \kappa (\Delta_1 + \Delta_2) + \frac{D_0}{2} (\nabla_1 + \nabla_2)^2 
    - \bigl( \mD(0) - \mD(\lvec) \bigr) : \nabla_1 \otimes \nabla_2.
\eeq
In the stationary state, and assuming statistical homogeneity and isotropy for the scalar covariance, \emph{e.g.} $C_2(\xvec_1,\xvec_1+\lvec,t) = C_2(\ell)$, this becomes
\beqs\label{eq:hopf_2}
\beq
    \mM_2^{\circ} C_2(\ell) = -\Phi(\ell), 
\eeq
\beq 
    \mM_2^{\circ}(\kappa) 
    =
    2\kappa \ell^{-1} \pd_\ell \left( \ell \pd_\ell \right) 
    + \left( D_0 - \mD_\parallel \right) \pd_{\ell\ell}^2 
    + \left( D_0 - \mD_\perp \right) \ell^{-1} \pd_{\ell}.
\eeq
\eeqs
In the rough regime $\eta \ll \ell \ll 1$ of Eq.~\eqref{eq:roughdecay}, the operator simplifies to 
\beq\label{eq:roughM2}
    \mM_2^{\circ}(\kappa) 
    \approx 
    \ell^{-1} \pd_\ell \left( \left( 2\kappa + \frac{D_1}{1+h} \ell^{2h} \right) \ell \pd_\ell \right),
\eeq
which yields the scaling laws
\beq\label{eq:scalingtheta}
    S^\theta_2(\ell) :
    =  
    2 \left( C_2(0) - C_2(\ell) \right)
    \sim 
    \begin{cases}
        & \dfrac{\varepsilon_\theta}{2\kappa} \ell^2, \quad \ell\ll\eta_\theta \\
        & \dfrac{\varepsilon_\theta}{D_1} \dfrac{1+h}{1-h} \ell^{2-2h}, \quad \ell\gg\eta_\theta,
    \end{cases}
\eeq
The scaling relations \eqref{eq:scalingtheta} feature the scalar dissipative scale 
\beq
    \eta_\theta = \left( \dfrac{2\kappa}{D_1} \right)^{1/2h}.
\eeq
In our computational setup later described in Section~\ref{sec:setup}, we choose the diffusivity as
\beq\label{eq:kappaD1}
   \kappa \propto D_1 \eta^{2h}.
\eeq
This ensures that $\eta_\theta \simeq \eta$, making the inertial range of scalar and velocity coincide. This is akin to considering Schmidt numbers of order unity \citep{donzis2014turbulent}. In the smooth regime  $\ell \ll \eta \ll 1$ of Eq.~\eqref{eq:smoothdecay}, the two-particle operator is dominated by its diffusive part, that is $\mM_2^{\circ} \simeq 2\kappa \ell^{-1} \pd_\ell \left( \ell \pd_\ell \right)$ and this again yields the asymptotics $\ell \ll \eta_\theta$ featured in Eq.~\eqref{eq:scalingtheta}. These asymptotics  imply anomalous scalar dissipation in the stationary state, through the identity $\varepsilon_\theta  =  \lim_{\kappa \to 0} \lim_{\xvec_2 \to \xvec_1} \kappa\, \Exp \left[ \nabla_1 \theta(\xvec_1) \cdot \nabla_2 \theta(\xvec_2) \right]$.

\subsubsection{Flatness anomaly}

Dimensional analysis would suggest a monofractal scaling in the inertial range as
\beq\label{eq:normal_scaling}
    S_p^\theta(\ell)
    =
    \Exp \left[ |\delta_\ell\theta|^p \right] \propto \ell^{p(1-h)}.
\eeq
This prediction proves correct for $p=2$ (see Eq.~\eqref{eq:scalingtheta}) but it is violated for higher orders, with the scalar exhibiting anomalous scaling in the form
\beq
    S_p^\theta(\ell)
    \sim
    \ell^{\zeta_p^\theta}, \qquad
    \zeta_p^\theta
    \neq
    p(1-h) \quad
    \text{for}\ p>2 .
\eeq
In the Kraichnan model, anomalous scaling comes from  the operator $\mM_n$ having non-trivial homogeneous isotropic zero-modes $Z_n$ satisfying 
\beq\label{eq:zero_mode}
    \mM^\circ_n Z_n = 0,
\eeq
which prescribe the inertial-range behavior of the correlation functions, and relate to statistical Lagrangian conservation laws \citep{gawedzki1995anomalous,falkovich2001particles}. The zero-mode exponents prove however cumbersome to extract in full generality. In practice, they are available only through perturbative computations. Estimates for the anomalous correction of the flatness exponent
\beq\label{eq:delta4}
    \Delta_4(h) = 2\zeta_2^\theta - \zeta_4^\theta, 
\eeq
can be obtained in limiting regimes. Near the discontinuous-flow limit, they take the form \citep{gawedzki1995anomalous, adzhemyan1998renormalization,adzhemyan2001anomalous, bernard1998slow}
\beq
    \Delta_4 (h) \sim 2h + 0.92624\, h^2 + O(h^3), \quad h \to 0.
\eeq
In the Batchelor limit $h \to 1$, estimates are not as crisp, but one expects perturbation series  in powers of $\gamma= (1-h)^{1/2}$ \citep{pumir1997perturbation,frisch1999lagrangian} leading to  
\beq
    \Delta_4 (h) \sim (4-2a)(1-h) - 2^{3/2}b (1-h)^{3/2} + O\left((1-h)^2\right), \quad h \to 1,
\eeq
where Lagrangian estimates of \citet{frisch1999lagrangian} suggest $a \simeq 0.06$ and $b \simeq 1.13$. We refer the reader to \citet{antonov2006renormalization} for a comprehensive review of the renormalization group approach to anomalous scaling in passive advection models.


\section{Computational setup}\label{sec:setup}

This section describes the finite-dimensional numerical realization of the Kraichnan model, specifying   the construction of the random flow on the torus, the stochastic time integration and the pseudo-spectral discretization, and our choices of physical parameters.

\subsection{Numerical implementation on the torus}

All simulations are performed on the two-dimensional torus $\mb{T}^2 = [0,2\pi]^2$. The generalized velocity and scalar fields are discretized on a uniform grid with $N^2$ points. The generalized velocity field is constructed as a finite-dimensional approximation of the Kraichnan velocity field \eqref{eq:flow_fourier}, replacing the integral representation by a finite sum over integer wavenumbers $\kvec \in \mb{Z}^2$. At each timestep, the velocity increments are  sampled over the time interval $[t_n,t_{n+1}]$ as
\beq\label{eq:flow_fourier}
    \Delta \Uvec_n (\xvec)
    =
    \sqrt{\dfrac{D_0}{Z}} \sum_{0 < |\kvec| \le N/3}\;
    e^{\iunit\kvec\cdot\xvec}\ \dfrac{g_\eta(|\kvec|)}{(|\kvec|^2+1)^{1/2+h/2}}\,
    \wh{\kvec}_\perp \mu(\kvec)\sqrt{ \Delta t}, 
\eeq
where $\Delta t = t_{n+1}-t_n$, and the complex random variables $\mu(\kvec) = \bigl( \mu_x(\kvec)+ i \mu_y(\kvec) \bigr) / 2$ are sampled from the centered Gaussian variables $\mu_l(\kvec)$ with covariance $\Exp \left[ \mu_l(\kvec_1) {\ol{\mu}}_m(\kvec_2) \right] = \delta_{lm}\, \delta(\kvec_1 - \kvec_2)$. The mode $\kvec = 0$ is excluded from the summation, ensuring that the velocity field has zero spatial mean almost surely. As a small-scale regularization, we use the gaussian filter
\beq 
    g_\eta(\kvec) = \exp \left[ -\frac{1}{2} \left( \frac{|\kvec|}{k_\eta} \right)^2 \right],
\eeq
where $k_\eta=2\pi/\eta$ is the regularizing (Kolmogorov) wavenumber. The normalization factor 
\beq\label{eq:discretenorm}
    Z = \sum_{\kvec \neq 0} \dfrac{g_\eta(|\kvec|)}{(|\kvec|^2+1)^{1+h}}
\eeq  
ensures that the total variance of the random flow increment is fixed by the parameter $D_0$, that is $\Exp \left[ \Delta {\Uvec}_n(\xvec) \cdot \Delta \Uvec_n(\xvec) \right] = 2D_0 \Delta t$ as in \S~\ref{sec:background}.

Similarly, the additive forcing increments \eqref{eq:forcing} are also discretized into random Fourier series as
\beq\label{eq:forcingtorus}
   \Delta F_n (\xvec)
    =
  \dfrac{1}{\sqrt{\tilde Z}} \sum_{0 < |\kvec| \le N/3}\,
  e^{\iunit\kvec\cdot\xvec} \phi(\kvec)\,
  \tilde \mu(\kvec) \sqrt{\Delta t},
\eeq
using the Gaussian spectral kernel \eqref{eq:forcing_kernel}, characterized by the scalar injection rate $ \varepsilon_\theta$ and forcing wavenumber $k_f$. The extra-normalization factor $\tilde Z = \sum_{\kvec \neq 0} |\phi(\kvec|^2 / \sqrt{\varepsilon_\theta}$ compensates the discretization effects, in order to ensure the energy injection rate $\epsilon_\theta = \Exp \Delta F_t(\xvec)^2 / \Delta t$. The random variables $\tilde \mu(\kvec) \overset{law}{\sim} \mu(\kvec)$ are sampled independently at each timestep.

\subsection{Pseudo-spectral discretization and time integration}

Restricting the summation to wavenumbers $0 < |\kvec| \le N/3$ in Eqs.~\eqref{eq:flow_fourier} and \eqref{eq:forcingtorus} is consistent with the standard 2/3-dealiasing rule used in pseudo-spectral methods. Spatial derivatives are computed in Fourier space, while products are evaluated in physical space, transformed back to Fourier space, and projected onto the modes $|\kvec| \le N/3$ \citep{orszag1971elimination, canuto2006spectral}. Time integration is performed using explicit Runge--Kutta (RK) schemes adapted to the Stratonovich formulation of Eq.~\eqref{eq:kraichnan_strat}. During each time step, a single realization of the random flow increment $\Delta\Uvec_n$ is sampled according to Eq.~\eqref{eq:flow_fourier} and kept fixed throughout all RK stages. This frozen-noise convention corresponds to the standard stage-wise interpretation of Stratonovich transport discretizations, and is consistent with the stochastic Heun and strong-stability-preserving Runge--Kutta (SSPRK3) schemes developed for transport-noise SPDEs~\citep{garcia2011comparison,cotter2019numerically}.

The present work relies on two stochastic time-integration schemes: a frozen-noise RK4 scheme used for all production runs, and an exponential RK4 variant employed for robustness checks.

\subsubsection{Frozen-noise RK4 scheme}

Let $\mathcal{P}_{2/3}$ denote the spectral projection onto the dealiased modes,
\[
   \mc{P}_{2/3} f = \sum_{|\kvec| \le N/3} f_{\kvec} e^{i\kvec\cdot\xvec}, \qquad
   f_{\kvec} = \frac{1}{(2\pi)^2} \int_{\mb{T}^2} f(\xvec) e^{-i\kvec\cdot\xvec}\,\df\xvec,
\]
and define the transport--diffusion operator
\[
   \mc{K}_n(\theta) 
   = 
   -\mc{P}_{2/3} \bigl( \Delta \Uvec_n \cdot \del \theta \bigr) + \kappa \Delta \theta\, \Delta t,
\]
where the same realization of the random flow increment $\Delta\Uvec_n$ is used throughout the RK stages. The additive forcing increment over one time step is denoted by $\Delta F_n$ and is sampled according to Eq.~\eqref{eq:forcingtorus}.

The RK4 update reads
\beq
   \theta^{n+1} = \theta^n + \frac16 (k_1+2k_2+2k_3+k_4) + \Delta F_n,
\eeq
with
\beqs
\begin{align}
   k_1 &= \mc{K}_n (\theta^n), \\ 
   k_2 &= \mc{K}_n \left( \theta^n + \frac{k_1}{2} \right), \\
   k_3 &= \mc{K}_n \left( \theta^n + \frac{k_2}{2} \right), \\
   k_4 &= \mc{K}_n \left( \theta^n + k_3 \right).
\end{align}
\eeqs
Since the same realization of $\Delta\Uvec_n$ is used in all four stages, this method should be interpreted as a frozen-noise RK discretization of the Stratonovich transport equation, rather than as a classical RK4 method applied to the full SPDE.

\subsubsection{Exponential RK4 variant}

The exponential RK4 scheme integrates the linear diffusion operator exactly through half-step integrating factors, while the advection is advanced using the same frozen-noise RK4 strategy. Owing to the additive nature of the forcing, its contribution is incorporated through the exact stochastic convolution associated with the integrating factor.
Let
\[
   E_{1/2} = \exp \left( \frac{\kappa\Delta\, \Delta t}{2} \right),
\]
and define the advection operator
\[
   \mc{A}_n(\theta) = -\mc{P}_{2/3} \bigl( \Delta \Uvec_n \cdot \del \theta \bigr).
\]
The forcing contribution over one time step is
\beq
   f_n(t) 
   = 
   \int_{t_n}^{t_{n+1}} \exp \left( (t_{n+1}-s) \kappa \Delta \right)\, \df F_s 
   = 
   \sum_{|\kvec| \le N/3} e^{i\kvec \cdot \xvec} \eta_n(\kvec),
\eeq
which is a centered Gaussian random field. Each of its Fourier modes has variance
\beq
   \Exp \left[ |{\eta}_n(\kvec)|^2 \right] 
   =
   |\phi(\kvec)|^2\, \frac{1-e^{-2\kappa|\kvec|^2\Delta t}}{2\kappa|\kvec|^2},
\eeq
with the continuous limit
\beq
   \Exp \left[ |{\eta}_n(\kvec)|^2 \right]
   =
   |{\phi}(\kvec)|^2 \Delta t \qquad 
   \text{as $\kappa|\kvec|^2 \to 0$.}
\eeq
The exponential RK4 update is
\beq
   \theta^{n+1} 
   =
   E_{1/2} \left[ \theta^{(0)} + \frac16 (a_1+2a_2+2a_3+a_4) \right] + f_n,
\eeq
where $\theta^{(0)} = E_{1/2}\theta^n$, and
\beqs
\begin{align}
   a_1 &= \mc{A}_n (\theta^{(0)}), \\
   a_2 &= \mc{A}_n \left( \theta^{(0)} + \frac{a_1}{2} \right), \\
   a_3 &= \mc{A}_n \left( \theta^{(0)} + \frac{a_2}{2} \right), \\
   a_4 &= \mc{A}_n \left( \theta^{(0)} + a_3 \right).
\end{align}
\eeqs

\subsection{Numerical parameters}

The parameters common to all simulations are summarized in Table~\ref{tab:common_params}.
\begin{table}
\centering
\begin{tabular}{ll}
\hline
Parameter & Value \\
\hline
Domain & $[0,2\pi]^2$ \\
Amplitude of the generalized velocity field & $D_0=1$ \\
Forcing scale & $k_f=2$ \\
Scalar injection rate & $\varepsilon_\theta=1$ \\
Regularization scale & $k_\eta=N/4$ \\
\hline
\end{tabular}
\caption{Common parameters used in the numerical simulations.}
\label{tab:common_params}
\end{table}
The scalar diffusivity is tuned depending on both the roughness parameter $h$ and the numerical resolution. It is parametrized as 
\beq
    \kappa = C_\kappa \dfrac{D_0}{Z} k_\eta^{-2h},
\eeq
where $Z$ is the discrete normalization factor defined in Eq.~\eqref{eq:discretenorm}, $D_0$ is the amplitude of the generalized field, $k_\eta$ is the regularizing wavenumber, and $C_\kappa$ is a dimensionless coefficient. In principle, in accordance with the prescription of Eq.~\eqref{eq:kappaD1}, we should choose $C_\kappa$ such that  $C_\kappa D_0/Z = O(D_1)$ where $D_1$ controls the decay of the generalized field in the rough regime; see Eq.~\eqref{eq:roughdecay}. In practice, though, $C_\kappa$ is tuned empirically as a function of $h$ to stabilize the small-scale scalar statistics while preserving an intermediate scaling range. For $h \le 1/2$, we report results with $C_\kappa D_0/Z \simeq D_1/3$. For larger values of $h$, we find that the numerical diffusion introduced by the time-integration scheme dominates any prescribed scalar diffusivity. We therefore simply set $\kappa=0$ in these cases. Table~\ref{tab:h_params} summarized the parameters used in the simulations.
\begin{table}
\centering
\begin{tabular}{lccccccccc}
\hline
$h$
& 0.1 & 0.2 & 0.33 & 0.4 & 0.5 & 0.6 & 0.67 & 0.8 & 0.9 \\
\hline
$C_\kappa$
& 6.5 & 4 & 2.7 & 2 & 1 & 0 & 0 & 0 & 0 \\
$D_0C_\kappa/Z$
& 0.31 & 0.30 & 0.32 & 0.29 & 0.19 & 0 & 0 & 0 & 0 \\
$D_1$
& 0.98 & 0.96 & 0.96 & 0.96 & 1 & 1.1 & 1.2 & 1.6 & 2.8 \\
\hline
\end{tabular}
\caption{
Roughness-dependent numerical parameters. The value of $Z$, defined in
Eq.~\eqref{eq:discretenorm}, is computed using $N=1024$ modes and the sharp
cutoff $g_\eta(\kvec)=1_{\kvec<N/3}$, but is essentially independent of the
resolution. The coefficient $D_1$ is defined in Eq.~\eqref{eq:roughdecay}.
}
\label{tab:h_params}
\end{table}
Finally, the resolution-dependent numerical parameters are summarized in Table~\ref{tab:resolution_params}. These values are used for the convergence tests reported in Section~\ref{sec:finiteband}. The $2/3$-rule dealiasing discards all the modes satisfying $|\kvec| > N/3$; the maximum effective wavenumber is therefore $k_{\max}=N/3$.
\begin{table}
\centering
\begin{tabular}{lccc}
\hline
Resolution $N^2$ & $k_{\max}$ & $k_\eta$ & $\Delta t$ \\
\hline
$512^2$  & $170$ & $128$ & $5\times10^{-5}$ \\
$1024^2$ & $341$ & $256$ & $10^{-5}$ \\
$2048^2$ & $682$ & $512$ & $10^{-6}$ \\
\hline
\end{tabular}
\caption{Resolution-dependent numerical parameters used in the convergence tests.}
\label{tab:resolution_params}
\end{table}

\subsection{Qualitative overview}

\begin{figure}
    \centering
    \includegraphics[width=\linewidth]{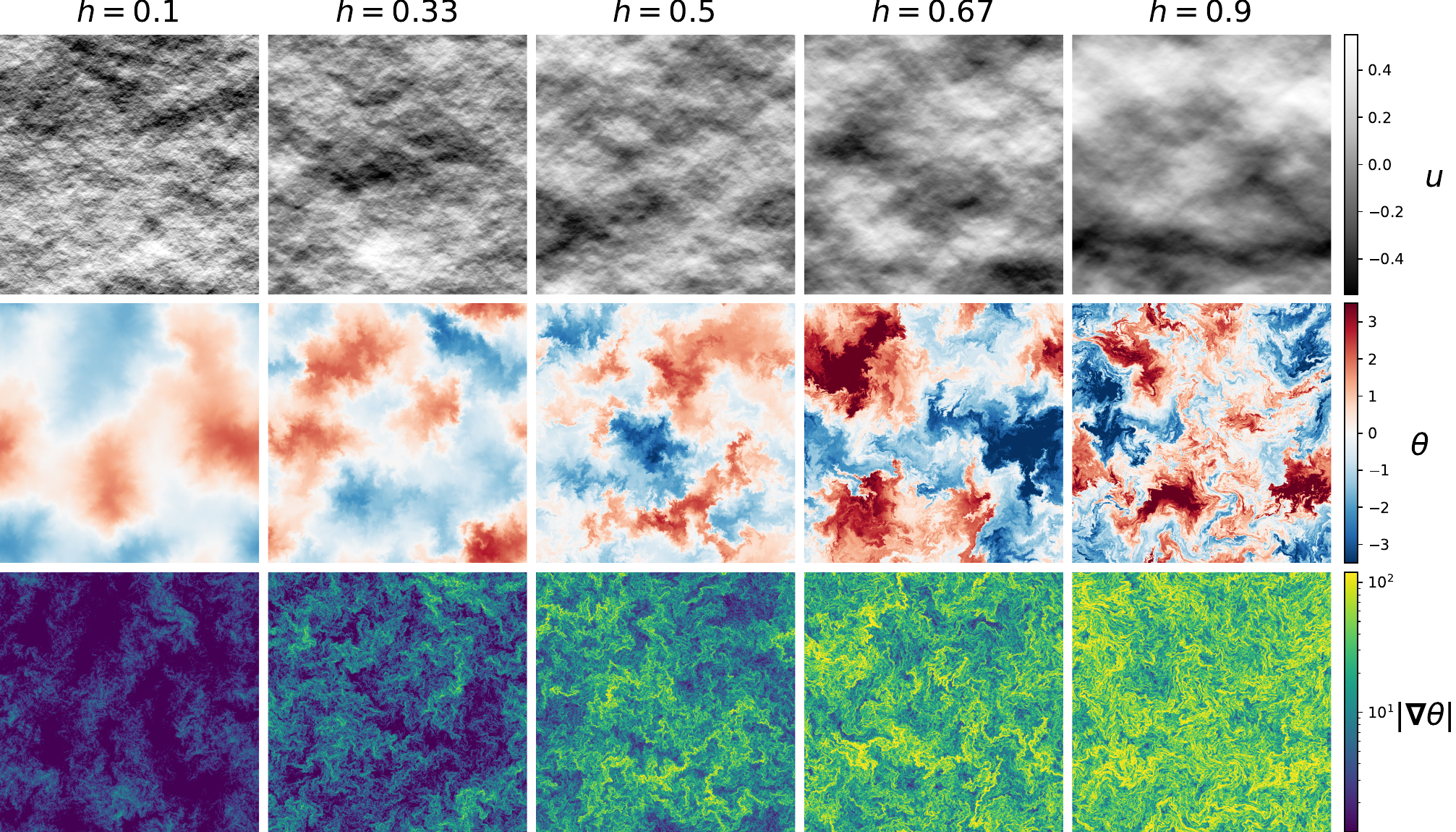}
    \caption{
    Sample snapshots at time $t=100$ for representative roughness exponents $h$. Top row: one component of the random flow $u$. Middle row: passive scalar $\theta$. Bottom row: scalar-gradient amplitude $|\del\theta|$.
    }
    \label{fig:spot}
\end{figure}

Figure~\ref{fig:spot} displays representative snapshots of one component of the generalized velocity field $\uvec := \Delta \Uvec / \Delta t^{1/2}$, the passive scalar, and the scalar-gradient amplitude at the final time $t=100$, for representative roughness exponents ranging from the nearly-discontinuous to the smooth case. As $h$ increases, the velocity field becomes progressively smoother, with more coherent large-scale structures. By contrast, the scalar field develops sharper fronts and finer-scale filaments, which are also reflected in the increasingly intense and localized gradient structures. These snapshots provide a first qualitative indication that the regularity of the random flow and that of the transported scalar are dual to each other, with $\Delta \theta^2 \propto \ell^{2(1-h)}$ as expected from the inertial-range estimate of Eq.~\eqref{eq:scalingtheta}.


\section{Second-order statistics}\label{sec:second_order}

Here we investigate the second-order statistics of the random generalized velocity field and of the passive scalar. 

\subsection{Global quantities}

We first present the numerical results on global-in-space quantities, such as the $L^2$ norms of $\theta$ and $\del\theta$, for various values of the roughness exponent $h$. We then analyze isotropic spectra and second-order structure functions of $\Uvec$ and $\theta$, and compare the numerical results with the theoretical scaling predictions. We conclude by assessing the robustness of these diagnostics with respect to spatial resolution by comparing results obtained at $N=512$, $1024$, and $2048$. Unless otherwise stated, the results shown in this section are computed at resolution $N=2048$. 

\begin{figure}
    \centering
    \includegraphics[width=\linewidth]{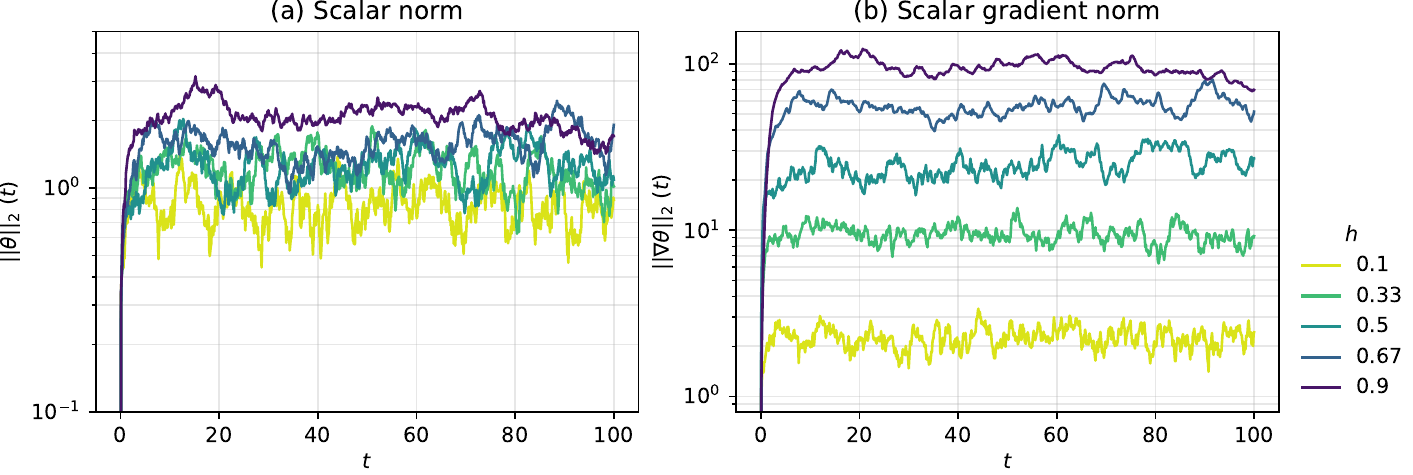}
    \caption{
    Time evolution of global scalar quantities for one pathwise realization and representative roughness exponents. (a) Spatial $L^2$ norm of the scalar $\theta$. (b) Spatial $L^2$ norm of the scalar gradient $\del\theta$. Both vertical axes are logarithmic.
    }
    \label{fig:norm}
\end{figure}

Figure~\ref{fig:norm}(a) shows the time evolution of the spatial $L^2$ norm of the scalar, while Figure~\ref{fig:norm}(b) shows the corresponding evolution of the $L^2$ norm of its gradient. Here and throughout, we use the normalized spatial $L^2$ norm, $\|\theta\|_{L_2} = \big( \frac{1}{(2\pi)^2} \int_{\mb{T}^2} \theta^2\, \dx \big)^{1/2}$. One can observe that, after a short transient, both quantities fluctuate around statistically stationary levels, with the amplitude of the scalar norm depending only moderately on $h$, whereas the amplitude of its gradient increases by several orders of magnitude when moving from rough to smooth flows. This quantitative trend confirms the qualitative picture of Figure~\ref{fig:spot}, where sharper fronts and finer-scale gradient structures are generated as $h$ increases. In what follows, all time-averaged statistics are evaluated over the interval $20 \leq t \leq100$, after discarding the initial transient.

\begin{figure}
    \centering
    \includegraphics[width=0.49\linewidth]{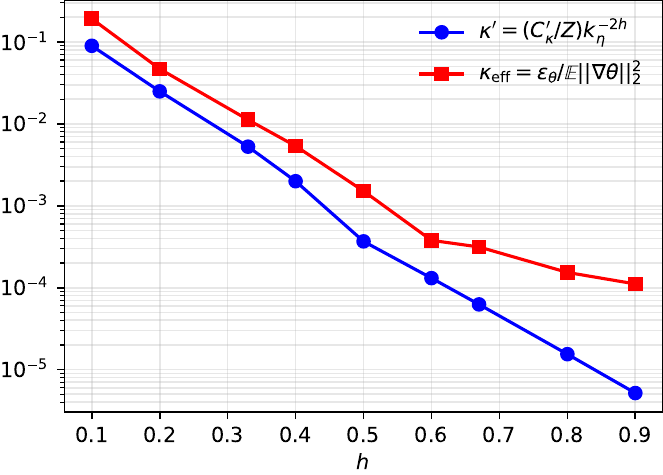}
    \caption{
    Comparison between the prescribed scalar diffusivity coefficient $\kappa' = (C_\kappa'/Z) k_\eta^{-2h}$ with $C_\kappa' = \max(1,C_\kappa)$ (blue circles) and the effective diffusivity $\kappa_{\mr{eff}} = \varepsilon_\theta/\Exp\|\del\theta\|_2^2$ (red squares) as functions of the flow roughness exponent $h$.
}
    \label{fig:effectivekappa}
\end{figure}

The saturation of the gradient allows us to estimate an effective numerical diffusivity as
\beq
    \kappa_{\mr{eff}} = \dfrac{\varepsilon_\theta}{\Exp \|\del \theta\|_{L_2}^2},
\eeq
shown in Figure \ref{fig:effectivekappa}. We find that $\kappa_{\mr{eff}}$ remains non-zero even for $h>0.5$, indicating that the time-integration scheme introduces an effective numerical dissipation allowing for a steady-state. The choice of setting $C_\kappa = 0$ in these runs is therefore effectively equivalent to choosing $C_\kappa<1$, as the explicit scalar diffusivity is then subdominant to the numerical one.

\subsection{Spectra and second-order structure functions}

\begin{figure}
    \centering
    \includegraphics[width=\linewidth]{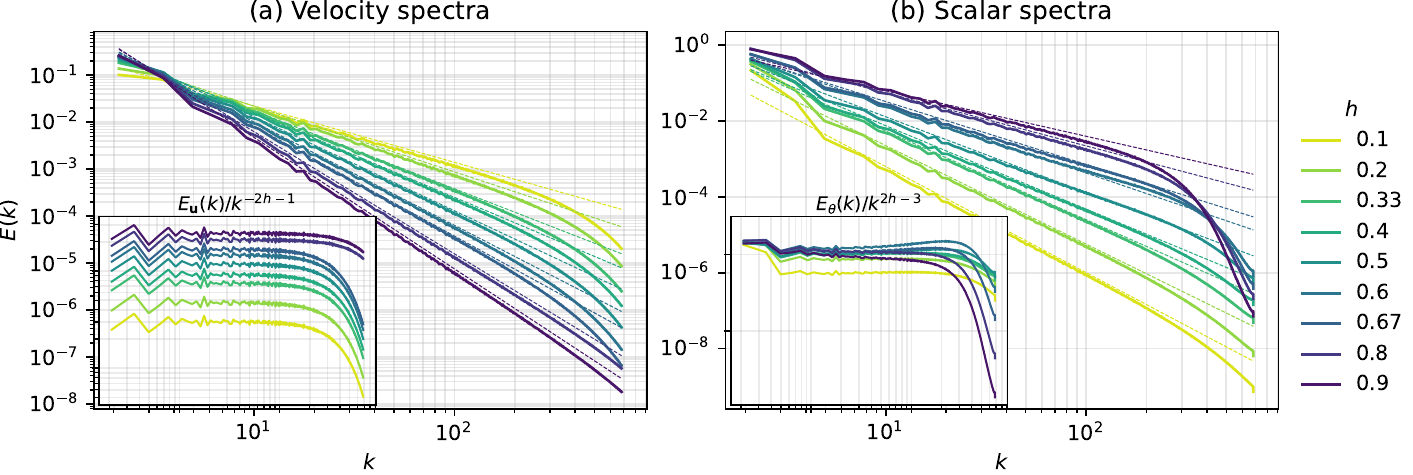}
    \caption{
    Time-averaged isotropic spectra as functions of the roughness exponent $h$. (a) Velocity spectra $E_{\uvec}(k)$, with dashed lines indicating the prescribed scalings $E_{\uvec}(k)\propto k^{-2h-1}$. (b) Scalar spectra $E_\theta(k)$, with dashed lines indicating the predicted scalings $E_\theta(k)\propto k^{2h-3}$. Insets show the spectra compensated by the corresponding theoretical power laws.
    }
    \label{fig:spec}
\end{figure}

We next turn to scale-dependent second-order statistics. Figure~\ref{fig:spec} shows the time-averaged isotropic spectra as functions of the roughness exponent $h$, together with the corresponding compensated spectra in the insets. Figure~\ref{fig:spec}(a) shows that the flow spectra reproduce the prescribed scaling $E_{\uvec}(k)\propto k^{-2h-1}$ over a broad range of scales, as expected from the spectral construction of the random velocity field.

Figure~\ref{fig:spec}(b), on the other hand, displays the corresponding scalar spectra. As $h$ increases, the scalar spectrum becomes less steep and contains relatively more energy at high wavenumbers. This is again consistent with the sharper scalar fronts showcased in the snapshots in Figure~\ref{fig:spot} and with the increase of the scalar-gradient norm shown in Figure~\ref{fig:norm}(b). The scalar spectra indicate that the predicted scaling $E_\theta(k)\propto k^{2h-3}$ is reasonably recovered over an intermediate range of wavenumbers from small values of $h$ up to approximately $h=2/3$. Near the smooth-flow limit ($h \to 1$), the spectra exhibit a pronounced high-wavenumber roll-off---a signature of the finite Fourier mode representation of the velocity field propagating into the scalar spectrum through the transport dynamics. By contrast, near the rough-flow limit ($h\to0$), the scalar spectra display the predicted behavior over an intermediate range when enough diffusion is imposed, as specified in Table~\ref{tab:h_params}. We discuss the influence of finite-size effects on the scalar statistics in more detail later in Section~\ref{sec:finitesize}.

\begin{figure}
    \centering
    \includegraphics[width=\linewidth]{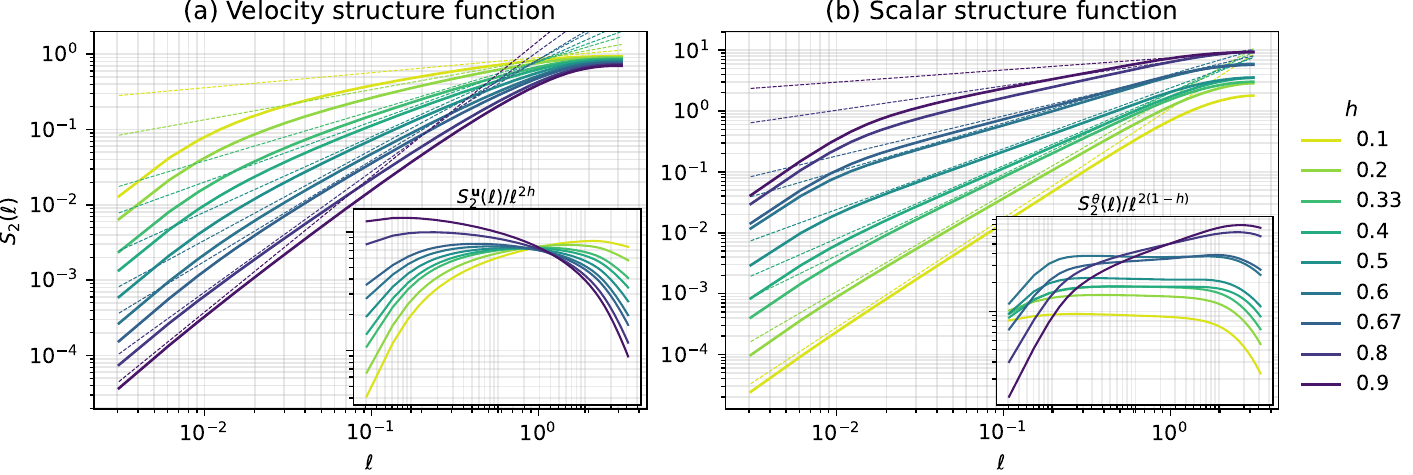}
    \caption{
    Second-order structure functions as functions of the roughness exponent $h$. (a) Longitudinal velocity structure functions $S_2^{\uvec}(\ell)$, with dashed lines indicating the predicted scalings $S_2^{\uvec}(\ell) \propto \ell^{2h}$. (b) Scalar structure functions $S_2^\theta(\ell)$, with dashed lines indicating the dual scaling behavior $S_2^\theta(\ell)\propto \ell^{2(1-h)}$. Insets show the structure functions compensated by the corresponding theoretical power laws.
    }
    \label{fig:sf2}
\end{figure}

Figure~\ref{fig:sf2} provides the corresponding physical-space characterization. Figure~\ref{fig:sf2}(a) reports the longitudinal second-order structure functions $S_2^{\uvec}(\ell)$ of the random velocity. While the prescribed power-law scaling is exact in Fourier space, its physical-space counterpart is affected by the finite spectral band available on the computational grid. As a result, the expected monofractal scaling behavior can only be recovered over a limited range of scales. The scaling $S_2^{\uvec}(\ell) \propto \ell^{2h}$ is most clearly observed for $1/3 \leq h \leq 2/3$, for which the compensated curves display a plateau over the inertial range. For smaller $h$, the plateau is displaced towards larger separations, whereas for larger $h$ it moves towards smaller scales. This indicates that larger scale separation, or equivalently more numerical resolution, would be needed to recover the theoretical scaling law over a wider range of spatial scales near the asymptotic regimes $h\to 0$ and $h\to 1$.

The scalar second-order structure functions are displayed in Figure~\ref{fig:sf2}(b). The theory of Kraichnan flows predicts a dual scaling behavior for the passive scalar, namely $S_2^\theta(\ell) \propto \ell^{2(1-h)}$ for separations in the inertial range; see Eq.~\eqref{eq:normal_scaling}. The numerical results corroborate this dual behavior: as $h$ increases, the scalar exponent $2(1-h)$ decreases, and the scalar structure functions become less steep. The compensated curves reveal an inertial range plateau for $h\leq 2/3$, indicating that the theoretical scaling law is recovered in this regime. Near the smooth-flow limit $h \to 1$, however, the plateau is strongly reduced and appears only over a narrow range close to the large-scale end of the resolved interval. This reflects the increasing roughness of the scalar field: smoother velocities produce rougher scalars, which require higher resolution to resolve. Consequently, the scalar statistics are strongly affected by the finite spectral representation of the velocity field when $h$ is large. By contrast, near the discontinuous-flow limit $h \to 0$, the theoretical scalar scaling can be observed once the dissipative scale is properly controlled.

\subsection{Effect of numerical resolution}

\begin{figure}
    \centering
    \includegraphics[width=\linewidth]{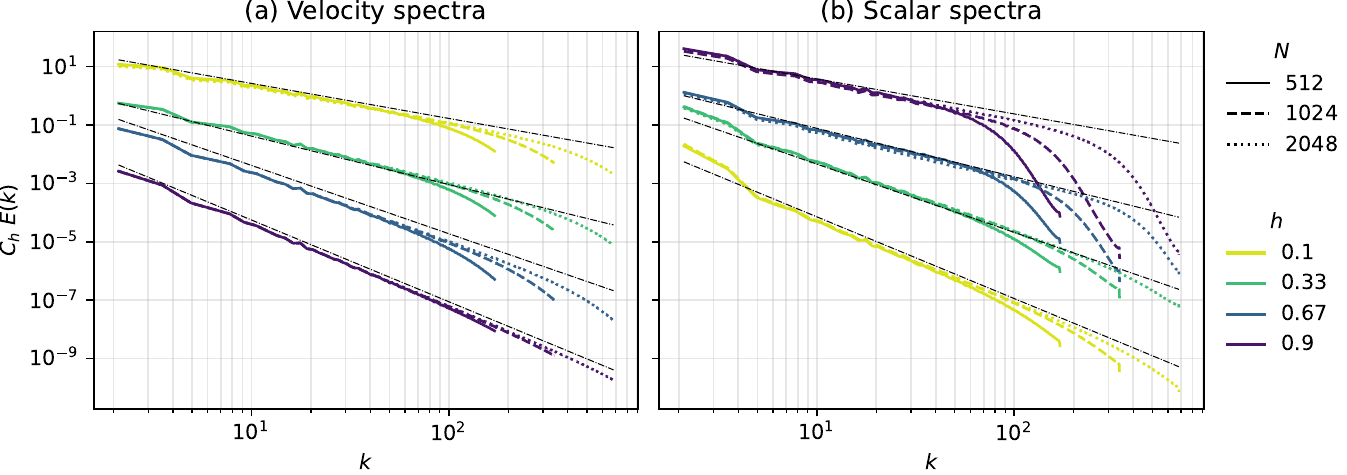}
    \caption{
    Resolution convergence of the time-averaged spectra. (a) Velocity spectra $E_{\uvec}(k)$. (b) Scalar spectra $E_\theta(k)$. The curves are shown for representative roughness exponents $h=0.1,0.33,0.67,0.9$ and resolutions $N=512,1024,2048$. A roughness-dependent factor $C_h$ is used only to separate the curves vertically. Black dash-dotted lines indicate the corresponding theoretical slopes.
    }
    \label{fig:conv_spec}
\end{figure}

We finally assess the robustness of the second-order statistics with respect to spatial resolution. Figure~\ref{fig:conv_spec} compares the time-averaged spectra obtained at $N=512$, $1024$, and $2048$, for representative roughness exponents spanning different regimes. For visual clarity, the curves are shifted vertically. For the random velocity field, shown in Figure~\ref{fig:conv_spec}(a), the spectra at different resolutions collapse well over their common resolved range. Increasing $N$ mainly shifts the high-wavenumber cutoff to larger values of $k$, thereby extending the range over which the prescribed spectral scaling is observed.

Figure~\ref{fig:conv_spec}(b) shows the corresponding scalar spectra, for which we obtain a similar convergent behavior: for small and intermediate values of $h$, the spectra at different resolutions agree well over their common resolved range, and increasing $N$ also extends the high-wavenumber range, similar to the velocity case. For $h=0.9$, the scalar spectra at different resolutions still collapse over their common resolved range, but the inertial range at a given resolution is narrower than for other values of $h$. Recovering a comparable extent of scaling range therefore requires higher resolution. This suggests that, for larger $h$, convergence of the scalar statistics with respect to resolution is slower than for smaller values of $h$.

\begin{figure}
    \centering
    \includegraphics[width=\linewidth]{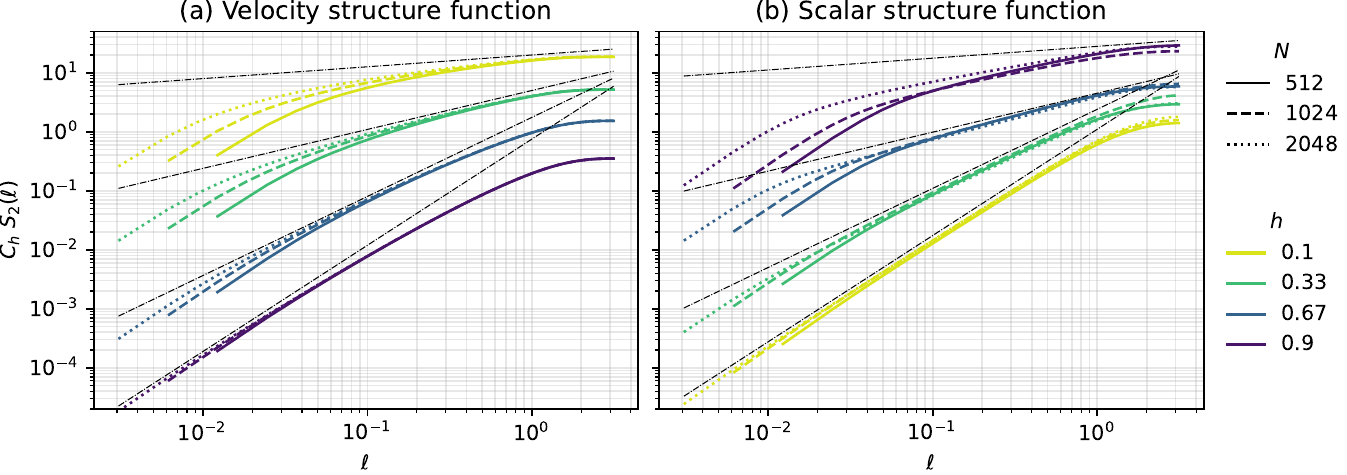}
    \caption{
    Resolution convergence of the second-order structure functions. (a) Longitudinal random-flow structure functions $S_2^{\parallel}(\ell)$. (b) Scalar structure functions $S_2^\theta(\ell)$. The curves are shown for representative roughness exponents $h=0.1,0.33,0.67,0.9$ and resolutions $N=512,1024,2048$. A roughness-dependent factor $C_h$ is used only to separate the curves vertically. Black dash-dotted lines indicate the corresponding theoretical slopes.
    }
    \label{fig:conv_sf2}
\end{figure}

We also conduct a similar convergence study in physical space. Figure~\ref{fig:conv_sf2}(a) reports the longitudinal second-order structure functions of the random velocity. The curves at different resolutions agree well over most of the common range of separations, while resolution effects are mainly visible at the smallest separations. These effects are more pronounced for small $h$, reflecting the difficulty of representing very rough random fields with a finite number of Fourier modes. For larger $h$, the flow structure functions are less sensitive to resolution at small separations.

Figure~\ref{fig:conv_sf2}(b) shows the scalar second-order structure functions. Here the resolution dependence displays the opposite tendency. For small and intermediate values of $h$, the curves collapse comparatively well over the intermediate range of separations, once the dissipative scale is properly controlled. For $h=0.9$, however, the scalar structure functions converge more slowly over the apparent inertial range: doubling the resolution does not bring the structure functions noticeably closer to the predicted scaling; they remain far from the theoretical behavior even at the highest resolution considered. This dual convergence behavior is consistent with the opposite dependence of the velocity and scalar regularities on $h$: rougher flows are harder to represent directly in physical space, whereas smoother flows generate rougher scalar fields whose inertial-range statistics require larger scale separation.

Overall, the second-order diagnostics show a consistent dual dependence on the roughness exponent. Increasing $h$ smooths the prescribed random flow, but roughens the transported scalar, leading to stronger gradients and enhanced high-wavenumber scalar content. The theoretical second-order scalar scaling is well recovered over an inertial range of scales for $h \leq 2/3$. Near the smooth-flow limit ($h\to1$), finite-size effects and limited scale separation become the dominant numerical limitations. These effects are analyzed in more detail in Section~\ref{sec:finitesize}.


\section{Characterization of intermittency}\label{sec:intermittency}

We now investigate the intermittent nature of the passive scalar through higher-order statistics. We first examine the scale dependence of the local flatness, along with the associated fourth-order anomaly $\Delta_4(h)$. We then show numerical results for the probability density functions (PDFs) of scalar increments, readily accessible within our Eulerian simulations, which provide a more complete description of the non-Gaussian character of the scalar fluctuations. We conclude by analyzing the higher-order scaling exponents $\zeta_p^\theta$ and their departure from log-normal behavior.

\subsection{Flatness anomaly}

\begin{figure}
    \centering
    \includegraphics[width=\linewidth]{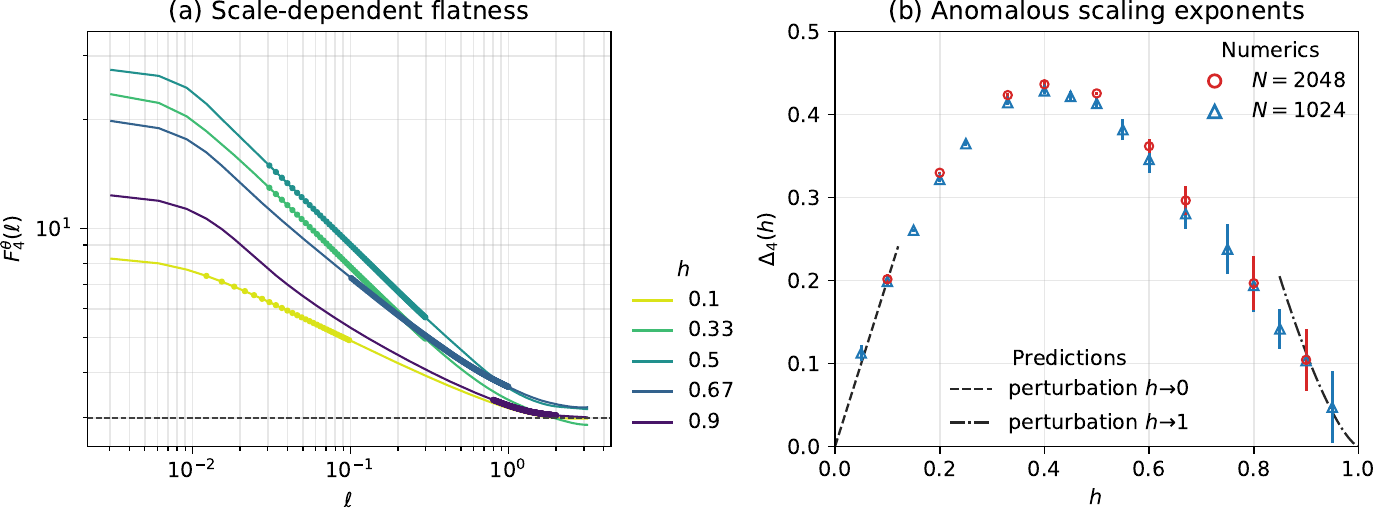}
    \caption{
    Fourth-order statistics of the passive scalar. (a) Local flatness $F_4^\theta(\ell)$ for representative roughness exponents $h$; the dashed horizontal line indicates the Gaussian reference value $F_4=3$. Markers indicate the nominal fitting windows used to estimate $\Delta_4(h)$. (b) Fourth-order anomalous correction $\Delta_4(h)$ as a function of $h$, for resolutions $N=1024$ and $N=2048$. Error bars are estimated by shifting the nominal fitting windows in log-space. Dashed and dash-dotted curves indicate perturbative predictions near the rough- and smooth-flow limits, respectively.
    }
    \label{fig:delta}
\end{figure}

Numerical results for the flatness $F_4^\theta(\ell)$ (see Eq.~\eqref{eq:flatness}) are presented in Figure~\ref{fig:delta}. Panel (a) reveals a systematic growth of the flatness towards smaller separations, for every value of $h$. More specifically, at large separations, the flatness approaches the Gaussian reference value $F_4=3$, whereas at small separations it becomes substantially larger, indicating that scalar increments become increasingly non-Gaussian at small scales. Equivalently, rare intense scalar fluctuations contribute more strongly to high-order moments than one would expect from the naive dimensional prediction---a hallmark of intermittency in turbulent flows. We also observe that the dependence on $h$ is not monotone. The growth of the flatness is more pronounced for intermediate values of $h$, while the roughest and smoothest cases display weaker deviations from Gaussianity over the resolved range.

In panel (b) we show results for the flatness exponent $\Delta_4(h)$, defined in Eq.~\eqref{eq:delta4}, as a function of $h$. We recall that a positive value of $\Delta_4(h)$ measures the departure from the self-similar relation $\zeta_4^\theta = 2\zeta_2^\theta$. The exponent $\Delta_4$ is estimated by fitting the local flatness according to $F_4^\theta(\ell)\sim \ell^{-\Delta_4}$ over an $h$-dependent nominal fitting window. These nominal windows are indicated by the markers in Figure~\ref{fig:delta}(a). They are chosen from the approximate plateau ranges observed in the compensated scalar second-order structure functions in Figure~\ref{fig:sf2}(b), so that the fourth-order fit is performed over the range where the second-order scalar scaling is best observed. To estimate the uncertainty associated with the choice of fitting interval, we repeat the fit over shifted windows obtained by multiplying the nominal bounds $(\ell_{\min},\ell_{\max})$ by factors logarithmically spaced between $0.75$ and $1.25$. The error bar is defined as half of the difference between the largest and smallest fitted values.

With this fitting strategy, the estimates obtained at resolutions $N=1024$ and $N=2048$ are in reasonable agreement over most of the range of $h$, indicating that the fourth-order anomaly is robust with respect to spatial resolution. The non-monotone behavior already apparent in panel (a) is also observed at the level of the flatness exponent: $\Delta_4(h)$ has a parabolic-like shape, with a maximum at intermediate values of $h$, and trends towards zero as $h$ approaches either $0$ or $1$. This is consistent with previous Lagrangian studies of the Kraichnan model \citep{frisch1999lagrangian,shraiman2000scalar,falkovich2001particles}. The dashed and dash-dotted curves indicate perturbative predictions near the rough- and smooth-flow limits. The numerical estimates agree well with the asymptotic behavior near both limits, while the maximum at intermediate $h$ reflects the strongest anomalous correction in the resolved range.

\subsection{Higher-order scaling anomalies}

\begin{figure}
    \centering
    \includegraphics[width=\linewidth]{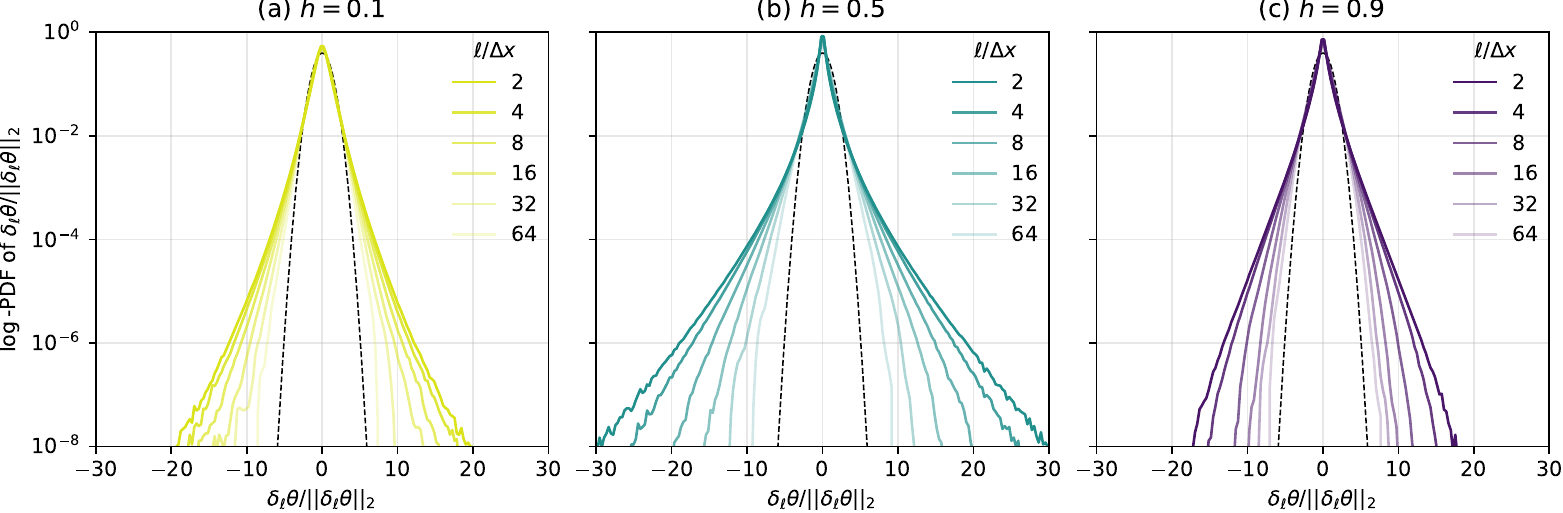}
    \caption{
    Probability density functions of normalized scalar increments $\delta_\ell\theta/\|\delta_\ell\theta\|_2$ for different separation scales. Results are shown for (a) $h=0.1$, (b) $h=0.5$, and (c) $h=0.9$, with separations $\ell/\Delta x=2,4,8,16,32,64$. The dashed black curve indicates the Gaussian reference. The vertical axes are logarithmic.
    }
    \label{fig:pdf}
\end{figure}

The non-Gaussian character of the scalar fluctuations is further illustrated by the PDFs of normalized scalar increments $\delta_\ell\theta/\|\delta_\ell\theta\|_2$. A useful advantage of the Eulerian simulations is that the full scalar field is available at each time, allowing scalar increments to be sampled over all grid points and over several separation scales. Figure~\ref{fig:pdf}(a)--(c) shows these PDFs for $h=0.1$, $0.5$, and $0.9$, and for different separations. For each value of $h$, the PDFs become increasingly non-Gaussian as $\ell$ decreases: the central peak sharpens and the tails broaden compared with the Gaussian reference. This scale dependence is consistent with the growth of local flatness at small separations in Figure~\ref{fig:delta}(a), and indicates that small-scale scalar increments are dominated by rare intense events associated with sharp fronts and localized gradient structures. The comparison across the three panels also shows that the strength of the tails depends non-monotonically on $h$. The intermediate case $h=0.5$ displays the broadest tails over the range of separations considered, whereas the tails are weaker in the rougher and smoother cases represented by $h=0.1$ and $h=0.9$. This agrees with the behavior of the fourth-order anomaly in Figure~\ref{fig:delta}(b), which reaches its maximum at intermediate values of $h$ and decreases toward both limiting regimes.

\begin{figure}
    \centering
    \includegraphics[width=\linewidth]{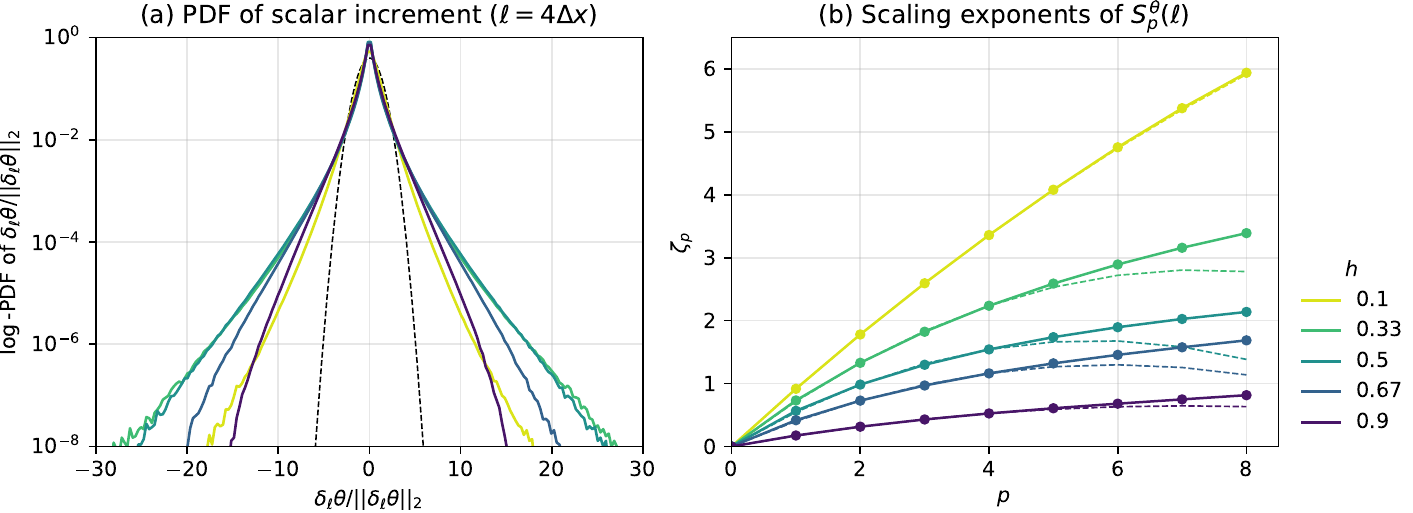}    
    \caption{
    (a) Probability density functions of normalized scalar increments $\delta_\ell\theta/\|\delta_\ell\theta\|_2$ at fixed separation $\ell=4\Delta x$, for different roughness exponents $h$; the dashed black curve indicates the Gaussian reference. (b) Estimated scaling exponents $\zeta_p^\theta$ as functions of the order $p$. Solid curves show the numerical estimates, and dashed curves show the log-normal-type fits defined in Equation~\eqref{eq:lognorm}.
    }
    \label{fig:zeta}
\end{figure}

Figure~\ref{fig:zeta}(a) gives a complementary comparison at fixed separation $\ell=4\Delta x$, allowing the dependence on $h$ to be seen more directly. The PDFs remain strongly non-Gaussian for all values of $h$, with sharp central peaks and heavy tails relative to the Gaussian reference. The broadest tails again occur for intermediate values of $h$, while the roughest and smoothest cases show weaker tails. This confirms that the reduction of the anomaly near the two limiting regimes is visible not only through the fitted value of $\Delta_4(h)$, but also at the level of the full increment distributions.

Figure~\ref{fig:zeta}(b) shows the higher-order scaling exponents $\zeta_p^\theta$, obtained from fits of the structure functions $S_p^\theta(\ell)\sim \ell^{\zeta_p^\theta}$ over the same $h$-dependent fitting windows used above. The exponents decrease with $h$, reflecting the progressive roughening of the scalar field: smoother random flows generate scalar fields with stronger gradients and smaller scaling exponents. The dependence of $\zeta_p^\theta$ on $p$ is nonlinear, especially at higher orders, which indicates a departure from self-similar scaling.
The dashed curves show log-normal-type fits of the form
\beq\label{eq:lognorm}
    \zeta_p^{\mr{LN}} = \alpha p + \beta p(p-2), \qquad
    \alpha = 1-h, \qquad
    \beta = \frac{\zeta_4^\theta-2\zeta_2^\theta}{8}.
\eeq
They provide a log-normal-type comparison constrained by the measured second- and fourth-order exponents, reasonably describing the overall curvature of $\zeta_p^\theta$ near the rough- and smooth-flow limits. For intermediate values of $h$, where the fourth-order anomaly is largest, the agreement deteriorates at higher orders: the measured exponents display weaker curvature and tend to lie above the log-normal curves. This suggests that higher-order scalar intermittency in the intermediate roughness regime is weaker than that captured by the log-normal comparison.

The good agreement observed near the two limiting regimes is to a large extent a consequence of the smallness of the anomaly itself. Since $\beta$ is proportional to $\Delta_4(h)$, which trends towards zero as $h$ approaches either $0$ or $1$, the quadratic contribution $\beta p(p-2)$ remains subdominant over the whole range of orders considered, and the log-normal prediction stays close to the linear behavior $\alpha p$. In this sense, the agreement is trivial, reflecting the weakness of intermittency in these regimes rather than the strict validity of the log-normal ansatz. This apparent agreement can also be understood in the light of the perturbative expansions of \citet{bernard1998slow} and \citet{pumir1997perturbation}. Both limits $h\to0$ and $h\to1$ correspond to solvable Gaussian points around which the anomalous exponents are computed perturbatively, with the leading correction to the linear behavior being quadratic in $p$. Near these points, a quadratic ansatz is therefore expected to capture the leading departure from self-similar scaling, irrespective of any log-normal assumption: any two-parameter quadratic form constrained by the measured $\Delta_4(h)$ would perform equally well. For intermediate values of $h$, by contrast, the flow is far from both Gaussian points, the expansion parameter is no longer small, and there is no reason for the quadratic term to dominate the higher-order corrections. The deterioration of the log-normal comparison in this regime is thus expected.

Physically, the deviation from log-normality at intermediate $h$ is consistent with the structures visible in the scalar snapshots of Figure~\ref{fig:spot}. Sharp fronts separating large regions of nearly uniform scalar---the ramp-and-cliff structures characteristic of scalar turbulence---make the high-order moments increasingly dominated by a small number of intense increments. Such structures are known to produce a tendency towards saturation of the scaling exponents at large $p$, rather than the downward curvature predicted by a quadratic ansatz. While the range of orders accessible here is too limited to establish saturation, the reduced growth of $\zeta_p^\theta$ observed at the largest values of $p$ is compatible with this scenario, and echoes the behavior reported for scalar turbulence in three-dimensional Navier--Stokes flows \citep{iyer2018steep}.


\section{Finite-size effects}\label{sec:finitesize}

The results of Section~\ref{sec:second_order} indicate that the main numerical limitations arise near the asymptotic rough- and smooth-flow regimes. In this section, we investigate these finite-size effects in more detail. We first consider the role of the scalar diffusivity in the discontinuous-flow limit, showing that an appropriate tuning of diffusivity improves the separation between inertial and dissipative scales. We then analyze the finite-band effects introduced by the truncated spectral representation of the random velocity and quantify their impact on the measured second-order statistics of the passive scalar.

\subsection{Tuning of diffusion coefficient}\label{sec:kappa}

\begin{figure}
    \centering
    \includegraphics[width=\linewidth]{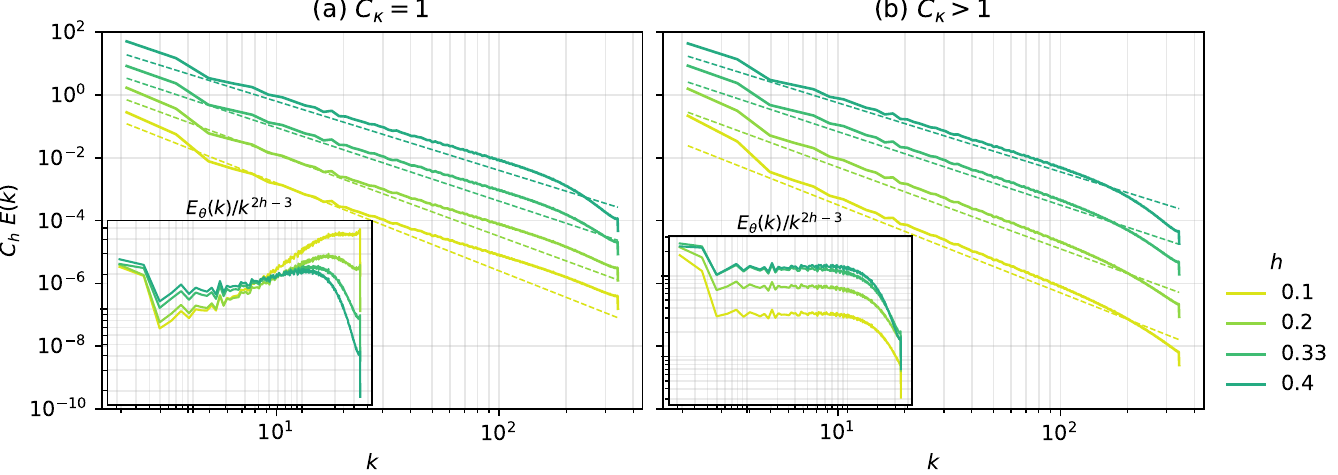}
    \caption{    
    Effect of scalar diffusivity for $h<1/2$. Scalar spectra are shown for $h=0.05,0.1,0.15,0.2$. (a) Baseline diffusivity coefficient $C_\kappa=1$. (b) Increased diffusivity coefficients $C_\kappa>1$, chosen according to Table~\ref{tab:h_params}. The curves are multiplied by a roughness-dependent factor $C_h$ only to separate them vertically. Dashed lines indicate the predicted scalings $E_\theta(k)\propto k^{2h-3}$, and insets show the corresponding compensated spectra.
    }
    \label{fig:spec_lim0}
\end{figure}

\begin{figure}
    \centering
    \includegraphics[width=\linewidth]{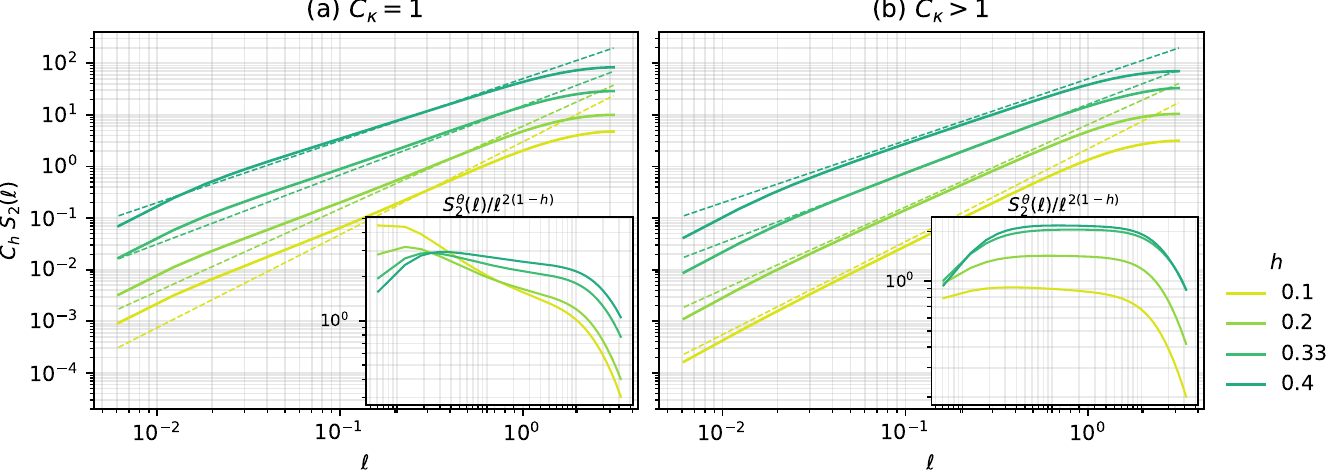}
    \caption{
    Effect of scalar diffusivity on second-order scalar structure functions for $h<1/2$. Results are shown for $h=0.05,0.1,0.15,0.2$. (a) Baseline diffusivity coefficient $C_\kappa=1$. (b) Increased diffusivity coefficients $C_\kappa>1$. The curves are multiplied by a roughness-dependent factor $C_h$ only to separate them vertically. Dashed lines indicate the predicted scalings $S_2^\theta(\ell)\propto \ell^{2(1-h)}$, and insets show the compensated structure functions.
    }
    \label{fig:sf2_lim0}
\end{figure}

To clarify the behavior in the discontinuous-flow limit (small-$h$), Figures~\ref{fig:spec_lim0} and~\ref{fig:sf2_lim0} compare simulations at resolution $N=1024$ with the baseline diffusivity coefficient $C_\kappa=1$ and with increased diffusivity coefficients $C_\kappa>1$, whose values are reported in Table~\ref{tab:h_params}. Figure~\ref{fig:spec_lim0} shows that, when $C_\kappa=1$, the compensated scalar spectra fail to develop a plateau and instead turn sharply upward at the highest wavenumbers, signaling a pile-up of scalar variance near the grid scale. Increasing $\kappa$ removes the spurious small-scale contributions and produces a clearer scaling range compatible with the theoretical prediction. The mechanism is clear: a larger $\kappa$ pushes the scalar dissipative scale $\eta_\theta$ back above the grid spacing, so that the scalar is regularized before the numerical cutoff acts. A similar tuning of the diffusivity as a function of $h$ was already required in the Lagrangian scheme of \citet{frisch1999lagrangian}.

The corresponding structure functions are shown in Figure~\ref{fig:sf2_lim0}. The same pattern appears in physical space: with insufficient diffusivity the compensated scalar structure functions show no satisfactory plateau, while increasing $\kappa$ flattens them over an intermediate range of separations. Thus, the small-$h$ limit can be made numerically more stable by tuning the diffusivity so that the dissipative scale remains resolved.

\subsection{Finite-band effects}\label{sec:finiteband}

The finite spectral support of the prescribed random velocity provides another source of finite-size effects. Even when the input spectrum follows the prescribed inertial-range scaling, the truncation to the resolved wavenumber interval $[k_{\min},k_{\max}]$ modifies the measured second-order statistics, particularly near the smooth-flow regime. To quantify this effect, we derive a finite-band correction for the velocity structure function and subsequently examine how these velocity distortions propagate to the passive scalar. Recall from Eq.~\eqref{eq:flow_covariance_par}, that ignoring discretization effects, the longitudinal second-order structure function $S_2^\parallel(\ell) = 2 \left(\mD^\parallel(0)-\mD^\parallel(\ell)\right)$ admits the spectral representation
\beq\label{eq:S2u_long_spectral}
    S_{2}^{\parallel}(\ell)
    =
    2\int_{k_{\min}}^{k_{\max}} E_{\uvec}(k) \left[ 1-J_0(k\ell)-J_2(k\ell) \right]\, \df k,
\eeq
where $E_{\uvec}(k)$ is the isotropic kinetic energy spectrum, which in the inertial range scales $E_{\uvec}(k) \propto k^{-1-2h}$. Using the change of variable $q=k\ell$, one factorizes Eq.~\eqref{eq:S2u_long_spectral} into 
\beq\label{eq:S2u_finiteband}
    S_{2}^{\parallel}(\ell)
    \propto
    \ell^{2h} I_{N,h}(\ell), \quad 
    I_{N,h}(\ell)
    :=
    \int_{k_{\min}\ell}^{k_{\max}\ell} \left[ 1-J_0(q)-J_2(q) \right] q^{-1-2h}\, \df q.
\eeq
involving the scaling part $\ell^{2h}$ and the finite-band correction factor $I_{N,h}$. The correction factor $I_{N,h}(\ell)$ is evaluated numerically by precomputing the cumulative integral in Eq.~\eqref{eq:S2u_finiteband} on a dense logarithmic grid in the similarity variable $q=k\ell$ and subsequently evaluating the correction through interpolation. 

The finite-band correction derived above accounts for the limited spectral support of the prescribed velocity field. The corresponding corrections for the passive scalar are deduced from stationary second-order Hopf equation derived in Eq.~\eqref{eq:hopf_2}. Using incompressibility, Eq.~\eqref{eq:hopf_2} can be written in the divergence form 
\beq\label{eq:scalar_divergence}
    \left[ \ell \pd_\ell \left( 2\kappa + \frac12 S_2^{\parallel}(\ell) \right) \pd_\ell C_2(\ell) \right]'
    =
    -\ell\Phi(\ell),
\eeq
where $C_2(\ell)$ denotes the scalar covariance and $\Phi(\ell)$ the forcing covariance. The quantity $2\kappa + \frac12 S_2^{\parallel}(\ell)$ thus acts as a scale-dependent effective (eddy) diffusivity governing the scalar statistics. Integrating Eq.~\eqref{eq:scalar_divergence} twice while imposing regularity at $\ell=0$ and using the Gaussian forcing covariance \eqref{eq:forcing_covariance} yields
\beq\label{eq:S2theta_pred}
    S_{2,\mr{pred}}^{\theta}(\ell)
    =
    \frac{4\varepsilon_\theta}{k_f^2} 
    \int_0^\ell \frac{1-\exp\left(-\frac{k_f^2r^2}{4}\right)}{\left(2\kappa + \frac12 S_2^{\parallel}(r)\right) r}\, \df r.
\eeq
Eq.~\eqref{eq:S2theta_pred} requires only the measured longitudinal velocity structure function as input, allowing for the investigation of how the finite-band distortions of the velocity field propagate to the second-order scalar statistics.

\begin{figure}
    \centering
    \includegraphics[width=\linewidth]{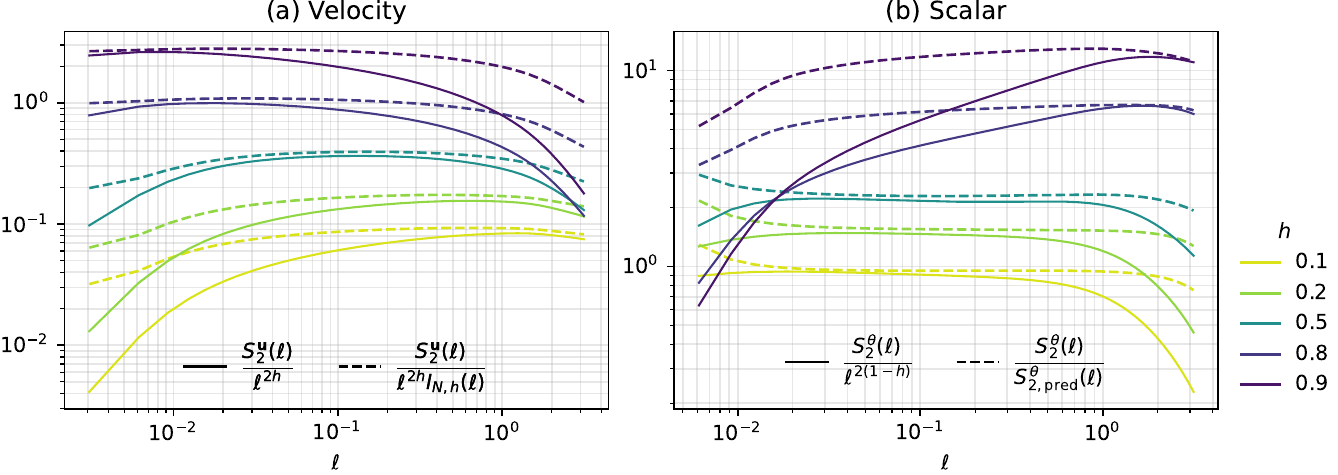}
    \caption{Diagnosis of finite-size effects in the velocity and scalar second-order structure functions. (a) Comparison between the standard inertial-range compensation $S_{2}^{\uvec}(\ell)/\ell^{2h}$ (solid lines) and the finite-band compensated quantity $S_{2}^{\uvec}(\ell)/[\ell^{2h}I_{N,h}(\ell)]$ (dashed lines) for representative roughness exponents $h$. (b) Comparison between the standard scalar compensation $S_{2}^{\theta}(\ell)/\ell^{2(1-h)}$ (solid lines) and the ratio $S_{2}^{\theta}(\ell)/S_{2,\mr{pred}}^{\theta}(\ell)$ (dashed lines), where $S_{2,\mr{pred}}^{\theta}$ is obtained from Eq.~\eqref{eq:S2theta_pred} using the measured longitudinal velocity structure function as input. In both panels, the curves are multiplied by roughness-dependent constants for visual clarity.
}
    \label{fig:S2_finite}
\end{figure}

In Figure~\ref{fig:S2_finite}, we show results for the second-order structure functions of velocity and passive scalar compensated by the corresponding finite-size predictions. Panel~(a) compares the conventional inertial-range compensation $S_{2}^{\uvec}(\ell)/\ell^{2h}$ with the finite-band compensated quantity $S_{2}^{\uvec}(\ell)/[\ell^{2h}I_{N,h}(\ell)]$. The latter is substantially flatter over a broader range of separations, demonstrating that much of the residual scale dependence originates from the finite spectral support of the prescribed velocity field. Panel~(b) investigates how these finite-band distortions propagate to the passive scalar. The solid curves show the infinite-size compensation $S_{2}^{\theta}(\ell)/\ell^{2(1-h)}$, whereas the dashed curves show the ratio $S_{2}^{\theta}(\ell)/S_{2,\mr{pred}}^{\theta}(\ell)$, where $S_{2,\mr{pred}}^{\theta}$ is obtained from Eq.~\eqref{eq:S2theta_pred} using the measured longitudinal velocity structure function as input. Compared with the power-law compensation, the velocity-conditioned prediction produces a substantially flatter behavior over a wide range of separations for all values of $h$. This indicates that the finite-size distortions observed in the scalar second-order statistics are largely inherited from the finite-resolution representation of the advecting velocity field through the closed second-order Hopf dynamics.


\section{Conclusion and perspectives}\label{sec:conclu}

We have developed a pseudo-spectral Eulerian framework for the numerical simulation of passive scalar transport in the Kraichnan model over the full range of admissible velocity roughness exponents. The proposed methodology combines a Fourier-space construction of the Gaussian random flow with frozen-noise Runge--Kutta time integration, providing direct access to complete scalar fields while remaining fully consistent with the white-in-time stochastic forcing. Compared with the classical Lagrangian approach \citep{frisch1999lagrangian}, the Eulerian formulation enables the direct computation of spatial statistics ranging from second-order structure functions to higher-order intermittency diagnostics.

The numerical results reproduce the main statistical predictions of the Kraichnan model across the entire range of roughness exponents. In particular, the simulations recover the expected dual behavior between the velocity and scalar regularities, the theoretical second-order scaling laws over their resolved inertial ranges, and the anomalous scaling associated with scalar intermittency. By combining inertial-range fitting guided by the second-order statistics with a systematic uncertainty estimation based on fitting-window shifts, the measured fourth-order anomalous corrections agree with the classical Lagrangian results of \citet{frisch1999lagrangian} throughout the interval $0<h<1$. The availability of full scalar fields further makes it possible to characterize intermittency through increment probability density functions and higher-order scaling exponents, revealing systematic departures from Gaussianity and from log-normal behavior, with the strongest deviations occurring near intermediate roughness exponents.

A central objective of this work was also to assess the capabilities and limitations of Eulerian simulations of the Kraichnan model. We showed that reliable simulations require different numerical treatments in the two asymptotic regimes. In the discontinuous-flow limit, the main numerical constraint is associated with the dissipative scale and can be controlled through an appropriate choice of the molecular diffusivity. In the smooth-flow limit, the principal limitation originates from the finite-band representation of the prescribed velocity field. By combining a finite-band correction for the velocity statistics with the second-order Hopf equation, we demonstrated that the corresponding distortions of the scalar statistics can largely be explained by the finite-resolution representation of the advecting flow. 

Beyond the classical Kraichnan model, the present work establishes a numerical foundation for studying passive scalar transport in more realistic stochastic velocity fields. In particular, recent multifractal extensions of Kraichnan flows incorporate intermittency directly into the velocity field through Gaussian multiplicative chaos. While these models have already been investigated for the statistics of the velocity field itself and for Lagrangian particle transport, passive scalar transport remains unexplored. The Eulerian framework developed here provides a natural platform for addressing this problem. The residual departures from the theoretical scaling laws associated to the finite spectral resolution also point to a specific direction for improvement: one should seek a numerical modelling of the velocity that does not rely on a truncated Fourier series, but instead resolves scales directly in physical space. In this regard, we conjecture that a wavelet-based construction of the velocity field would provide a more efficient representation of the carrier flow, reducing the deviations from the theoretical scaling laws due to finite-size effects. This perspective should carry over to the multifractal setting, where the Gaussian carrier is modulated by a multiplicative chaos.


\section*{Acknowledgements}

The authors are grateful to the Universit\'e C\^ote d'Azur's Center for High-Performance Computing (OPAL infrastructure) for providing resources and support. L.L. gratefully acknowledges Mireille Bossy for insightful discussions and suggestions.

\section*{Funding}

This work was supported by the French government through the France 2030 investment plan, managed by the French National Research Agency (ANR), as part of the Initiative of Excellence Universit\'e C\^ote d'Azur (grant number ANR-15-IDEX-01); the CAPES MATH-AmSud project Cha$^2$MAN (grant number AMSUD54247VM); the CNRS Emergence@Physique 2025 programme; and a Synergy grant from the Acad\'emie d'Excellence Syst\`emes Complexes at Universit\'e C\^ote d'Azur.

\section*{Competing interests}

The authors declare none.

\section*{Data availability}

The simulation code, analysis scripts and notebooks, and numerical data supporting the findings of this study are openly available at \url{https://doi.org/10.5281/zenodo.22145071} \citep{li2026dataset}.



\appendix
\section{Decay rates for the generalized velocity field}\label{sec:tauberian}

We here provide details on the smooh and rough decay rates prescribed by Eq.~\eqref{eq:roughdecay} and \eqref{eq:smoothdecay}. We start from Eq.~ \eqref{eq:flow_covariance} expressing the velocity correlation as 
\beqs\label{eq:flow_trace_covariance_rappel}
\beq
    \Exp \left[ \dU(\xvec) \cdot \dU(\xvec+\lvec) \right]
    =
    \dt\, \tr \mD (|\lvec|), 
\eeq
\beq
    \tr \mD (\ell)  
    =
    \dfrac{2\pi D_0}{Z} \int_{\mb{R}_+} \df k\, J_0(k\ell) \frac{k\, g_\eta^2(k)}{(k^2+1)^{1+h}},
\eeq
\eeqs
with the normalizing factor $Z(\eta,h)$ chosen such that 
\beq\label{eq:norm}
    \tr \mD (0) = 2D_0.
\eeq
Setting $\eta \to 0$ yields the explicit closed form
\beq\label{eq:Ih}
    \tr \mD (\ell) 
    = 
    \dfrac{2\pi D_0}{Z} \dfrac{\ell^h}{2^h\Gamma(1+h)} K_h(\ell),
\eeq
where $K_h$ is the modified Bessel function of the second kind \citep[see, e.g.,][]{abramowitz1948handbook}, with the small-$\ell$ expansion
\beq
    \ell^h K_h(\ell) 
    =
    2^{h-1} \Gamma(h) + 2^{-h-1} \Gamma(-h) \ell^{2h} + O(\ell^2).
\eeq
This yields 
\beq
   Z 
   = 
   \dfrac{\pi \Gamma(h)}{2 \Gamma(1+h)} \quad \text{and} \quad 
   \tr \mD (\ell)  
   \sim   
   2D_0 \left( 1+2^{-2h} \dfrac{\Gamma(-h)}{\Gamma(h)} \ell^{2h} \right),
\eeq
and prescribes the coefficient $D_1 = -D_0 2^{-2h} \frac{\Gamma(-h)}{\Gamma(h)}$ in Eq.~\eqref{eq:roughdecay}.

Now consider the smooth limit $\ell \to 0$ at fixed $\eta$ with an ultraviolet cutoff function of the form  $g_\eta(k) = g(k/k_\eta)$. Expanding the Bessel function as $J_0(k\ell) = 1-\dfrac{k^2\ell^2}{4} + \cdots$ yields the small-$\ell$ limit
\beqs
\beq
   \tr \mD (\ell)  
   = 
   \dfrac{2\pi D_0}{Z} \left( I_1(k_\eta) - I_2(k_\eta) \ell^2 + O(\ell^4) \right),
\eeq
\beq
   I_1(k_\eta)
   :=   
   k_\eta^{-2h} \int_0^\infty \dfrac{\df u\, u\, g(u)}{\left(u^2+k_\eta^{-2}\right)^{1+h}},
\eeq
\beq
   I_2(k_\eta)
   :=   
   \dfrac{k_\eta^{2-2h}}{4} \int_0^\infty \dfrac{\df u\, u^3\, g(u)}{\left(u^2+k_\eta^{-2}\right)^{1+h}}, 
\eeq
\eeqs
where the specific expressions of $I_1$ and $I_2$ come from the change of variables $k \to k/k_\eta$. The normalization \eqref{eq:norm} implies  
\beq\label{eq:Tr2}
   Z(k_\eta) = \pi I_1 \quad \text{and} \quad   
   \tr \mD (\ell) = 2D_0 \left(1 - \dfrac{I_2}{I_1}\ell^2) \right).
\eeq
Taking $k_\eta \to \infty$, we now compute
\beq\label{eq:I12}
   I_1(k_\eta) \sim \dfrac{1}{2h}, \qquad 
   I_2(k_\eta) \sim \dfrac{k_\eta^{2-2h}}{4} \int_0^\infty \df u\, u^{1-2h} g(u).
\eeq
In particular,
\beq\label{eq:I2}
   I_2 \sim 
   \begin{cases}
      & \dfrac{k_\eta^{2-2h}}{8(1-h)} \qquad \text{for the sharp cutoff } g(u):=1_{u<1} \\
      & \dfrac{k_\eta^{2-2h}}{8\Gamma(1-h)} \qquad \text{for the Gaussian cutoff } g(u):=e^{-u^2} \\
   \end{cases}.
\eeq
Combining Eqs.~\eqref{eq:Tr2}, \eqref{eq:I12} and \eqref{eq:I2} prescribes the coefficients
$D_2$ in Eq.~\eqref{eq:D2}.

\clearpage

\bibliographystyle{jfm}
\bibliography{num_kraichnan}

\end{document}